\documentclass[%
preprintnumbers,
nofootinbib,
 amsmath,amssymb,
 aps,
prd,
]{revtex4-2}

\usepackage{amssymb} % For the checkmark
\usepackage{graphicx,amssymb,slashed}
\usepackage{dsfont}
\usepackage{subcaption}% Includes subfigures
\usepackage[justification=raggedright]{caption}
\usepackage{dcolumn}% Align table columns on decimal point
\usepackage{bm}% bold math
\usepackage{hyperref}% add hypertext capabilities
\hypersetup{
  colorlinks   = true, %Colours links instead of ugly boxes
  urlcolor     = blue, %Colour for external hyperlinks
  linkcolor    = blue, %Colour of internal links
  citecolor   = red %Colour of citations
}

\usepackage{slashed}
\usepackage{parskip}

\def\be{\begin{equation}}
\def\ee{\end{equation}}
\def\bea{\begin{eqnarray}}
\def\eea{\end{eqnarray}}
\def\bear{\begin{array}}
\def\ear{\end{array}}
\def\bfig{\begin{figure}}
\def\efig{\end{figure}}
\def\bcen{\begin{center}}
\def\ecen{\end{center}}
\def\bi{\begin{itemize}}
\def\ei{\end{itemize}}

\usepackage{bm}% bold math
\usepackage{xcolor}%color texto
\usepackage{multirow}%tablas
\usepackage{slashed}%feynman slash con \slashed{D}
\usepackage{physics}
\usepackage{makecell}

\newcommand{\no}{\nonumber}

\newcommand{\mL}{\mathcal{L}}

\newcommand{\mO}{\mathcal{O}}

\newcommand{\fpi}{\mathring{F}_\pi}
\newcommand{\mpi}{M_{\pi}}
\newcommand{\md}{\mathring{m}_\Delta}
\newcommand{\m}{\mathring{m}}
\newcommand{\g}{\mathring{g}_A}
\newcommand{\pthree}{\mO(p^3)}
\newcommand{\pfour}{\mO(p^4)}
\newcommand{\pfive}{\mO(p^5)}
\newcommand{\chidof}{\chi^2/{\rm dof}}
\newcommand{\rad}{\langle r_1^{2}\rangle}
\newcommand{\radtwo}{\langle r_2^{2}\rangle}

\begin{document}

\title{Light quark mass dependence of nucleon electromagnetic form factors in dispersively modified chiral perturbation theory}

\author{Fernando Alvarado$^{1,2}$}
\email{Fernando.Alvarado@ific.uv.es}
\author{Di An$^2$}
\email{di.an@physics.uu.se}
\author{Luis Alvarez-Ruso$^1$}%
\email{Luis.Alvarez@ific.uv.es}
\author{Stefan Leupold$^2$}
\email{stefan.leupold@physics.uu.se}

\affiliation{$^1$ Instituto de F\'isica Corpuscular (IFIC), \\ Consejo Superior de Investigaciones Cient\'{i}ficas (CSIC) and Universidad de Valencia (UV)\\ E-46980 Paterna, Valencia, Spain}
\affiliation{$^2$ Institutionen f\"or fysik och astronomi, Uppsala universitet, Box 516, S-75120 Uppsala, Sweden}
\date{\today}

\begin{abstract}

The nucleon isovector electromagnetic form factors are calculated up to next-to-next-to-leading order 
by combining relativistic chiral perturbation theory (ChPT) of pion, nucleon, and $\Delta(1232)$ with dispersion theory. We specifically address the light-quark mass dependence of the form factors, achieving a good description of recent Lattice QCD results over a range of $Q^2\lesssim 0.6$ GeV$^2$ and $\mpi\lesssim 350$ MeV. For the Dirac form factor, the combination of ChPT and dispersion theory outperforms the pure dispersive and pure ChPT descriptions. For the Pauli form factor, the combined calculation leads to results comparable to the purely dispersive ones.  The anomalous magnetic moment and the Dirac and Pauli radii are extracted.
\end{abstract}

\maketitle

\section{\label{sec:intro} Introduction}

Elastic scattering of nucleons by point-like, leptonic probes is among the simplest observable processes sensitive to the nucleon’s internal structure~\cite{Gross:2022hyw}. Therefore, 
electromagnetic nucleon form factors (FFs) play a pivotal role in elucidating the intricate structure of nucleons and the underlying fundamental strong interaction governing their behaviour. There are many open questions concerning the nucleon electromagnetic FFs. To name a few: Does the electric FF of the proton have a zero crossing in the spacelike region \cite{Punjabi:2015bba,Gross:2022hyw}? Why does the neutron have a negative transverse charge density not only at the periphery but also at the centre \cite{Miller:2007uy,Carlson:2007xd}? In recent years the most significant question was probably the one related to the proton radius puzzle \cite{Pohl:2010zza}:  muonic hydrogen data contradicted the determination from ordinary (electronic) hydrogen and electron scattering. The extraction of the radius from $ep$ scattering led to different results depending on how to extrapolate to the photon point i.e.\ to zero photon virtuality $q^2 = - Q^2$, where the radius is determined~\cite{Belushkin:2006qa,Sick:2017aor,Alarcon:2017lhg}. 
Although several recent experimental results align with the muonic hydrogen measurement~\cite{Pohl:2010zza} and with determinations based on dispersion relations (see Fig.~5 of Ref.~\cite{Meissner:2022rsm} and references therein), the experimental scenario remains uncertain as apparent from Fig.~1 of Ref.~\cite{Xiong:2023zih}.

From the theoretical point of view, the ab-initio calculation of the nucleon FFs can be achieved using the Lattice QCD (LQCD) method. Although results are now available even at physical quark masses \cite{Djukanovic:2021cgp}, high-precision determinations are still hampered by large uncertainties when extrapolating the LQCD results to the photon point. Actually, at present, LQCD cannot rule out the original electronic hydrogen data even though the muonic hydrogen results are favored \cite{Djukanovic:2021cgp}.
Alongside the extrapolation to the photon point, the extraction of the nucleon electromagnetic radii involves the interpolation or extrapolation of the light-quark mass dependence of the FFs to the physical values. 

Effective field theory allows to undertake the challenge of predicting both the $Q^2$ and the quark mass dependencies of the FFs in a model independent way.

Furthermore, the study of the light quark mass dependence is interesting in its own right because it provides theoretical insight that might not be (easily) accessible from experimental data. Such a study will not only serve to understand the nucleon structure itself but also to test the basic properties of QCD: The interplay between explicit quark masses and the dynamical scale of QCD; chiral symmetries and the spontaneous symmetry breaking; long-range forces mediated by the emerging Goldstone bosons vs.\ standard short-range forces caused by confinement. 

Chiral perturbation theory (ChPT) \cite{Weinberg:1978kz,Gasser:1983yg} is the effective theory of strong interactions among massless pseudoscalar mesons, which emerge as the Goldstone bosons of the spontaneous chiral symmetry breaking. The structure of the interactions is fixed by symmetry (and symmetry breaking) considerations. The leading-order Lagrangian has only a small number of parameters. Systematic improvement by means of perturbation theory is possible as long as the typical four-momenta remain well below the symmetry-breaking scale ($p \ll \Lambda_\chi$), and at the price of introducing further low-energy constants (LECs). Baryons bring about the complication of a new scale, the baryon mass, which does not vanish in the chiral limit, where Goldstone bosons become massless. As a consequence, the power counting is disrupted~\cite{Gasser:1987rb}. Different procedures to systematically restore power counting, making perturbation theory feasible, have been developed (see for instance Ref.~\cite{Scherer:2012xha}). In the present study we adopt the extended on mass shell (EOMS) renormalisation scheme~\cite{Fuchs:2003qc}, preserving covariance and the analytic properties of loop amplitudes. By coupling the ChPT Lagrangian to external electroweak sources it is possible to study nucleon electromagnetic and axial form factors at low $Q^2$~\cite{Kubis:2000zd,Fuchs:2003ir,Schindler:2006it,Ledwig:2011cx,Bauer:2012pv,Flores-Mendieta:2015wir,Yao:2016vbz,HillerBlin:2017syu}. The explicit chiral symmetry breaking by light quark masses, from which pseudoscalar mesons acquire their masses, is naturally incorporated to the framework. This renders ChPT an ideal tool for the combined study of the $\mpi$ and $Q^2$ behaviour of nucleon FFs at $\mpi, \sqrt{Q^2} \ll \Lambda_\chi \sim 1\,$GeV~\cite{Ledwig:2011cx,Yao:2017fym,Djukanovic:2021cgp}.\footnote{In the isospin limit $m_u=m_d \equiv \hat{m}$, $M_\pi^2 = 2 B_0 \hat{m} + \mathcal{O}(p^4)$ so we indistinctly refer to the $\hat{m}$ or $M_\pi$ dependence.} 

By construction, ChPT is limited to the energy range where hadronic resonances are not excited. 

This limitation can become too restrictive if there are degrees of freedom which couple strongly to pions and nucleons and/or have relatively low masses. Phenomenologically, it is known that $\pi$-$N$ systems couple strongly to the $\Delta(1232)$ states. 
In addition, these states are relatively light: $\delta = m_\Delta - m_N \sim 300$~MeV.  
It is therefore advisable to include the $\Delta$ as a dynamical degree of freedom~\cite{Jenkins:1991es,Hemmert:1997ye,Pascalutsa:2002pi,Bernard:2003xf,Hacker:2005fh} to improve the convergence and extend the range of applicability of ChPT. In general, the study of nucleon properties: polarizabilities, couplings, form factors, in ChPT benefits from the dynamical treatment of the $\Delta$ resonance, as can be appreciated from this, by no means exhaustive, list of references~\cite{Yao:2016vbz,HillerBlin:2017syu,Thurmann:2020mog,Alvarado:2021ibw}. In different scenarios, different counting rules for $\delta$ are assumed. As will be discussed below, we follow the small-scale expansion~\cite{Hemmert:1997ye}, according to which $\mathcal{O}(\delta)\sim \mathcal{O}(p)$. 

Virtual photons couple to pion pairs in a p-wave state. On the other hand, it is well known that two pions in a p-wave state are strongly correlated by the $\rho$ meson. 
Though the coupling is also strong, this case is different from the $\Delta$ one,
where the momentum scales with another small parameter. 
The momentum of a pion resulting from the decay of an {\it on shell} $\Delta$ (i.e.\ with an invariant mass equal to the resonance Breit-Wigner mass) to $\pi N$ scales with the baryon mass difference $\delta$, which is a small parameter of the theory. In other words, such pion momenta can be regarded as small. The hard scale, the baryon mass, cannot make the pion momentum large,
because this hard scale remains with the nucleon, i.e.\ with the state that carries the conserved baryon number. 
In contrast, the decay of the $\rho$ meson into two pions provides a 
momentum to one or both of the pions, 
which should be considered large 
unless the $\rho$-meson mass, $m_\rho$, is regarded as a soft scale. 
In view of the fact that $m_\pi \ll m_\rho \sim \Lambda_\chi$,
the conservative approach is to regard the mass of the $\rho$ meson as a hard scale of the process. It is then difficult to include the $\rho$ meson as a dynamical degree of freedom in a low-energy effective field theory although this idea has been attempted, for instance in Refs.~\cite{Djukanovic:2009zn,Terschlusen:2012xw,Terschlusen:2016kje}.
In addition, there are plenty of phenomenological models to include the $\rho$ meson, often based on the concept of vector-meson dominance \cite{Gounaris:1968mw,Feynman:1973xc,Landsberg:1985gaz,Czerwinski:2012ry}. We refrain from the use of phenomenological models since we do not want to give up the key features of an effective field theory, being model independent, systematically improvable and having controlled uncertainties.
Therefore the task is to develop a systematic, model independent scheme that allows to resum terms $\sim p^2/m_\rho^2$, where $p$ denotes some typical pion momentum (or photon virtuality). 
In that way, it might  be possible to extend the range of applicability of ChPT concerning the expansions in momenta and/or in the pion mass. 

The $\rho$ meson is observed as an elastic resonance in pion-pion scattering. 
(Actually the same remark applies to the $\Delta$ as an 
elastic resonance in pion-nucleon scattering.) 
Its coupling to all other channels (virtual photon, but also three- and four-pion states \cite{Hanhart:2012wi}) is very weak. 
From the point of view of S-matrix theory with its 
focus on asymptotic states, 
this suggests that the $\rho$ meson can be included via the two-pion p-wave phase shift provided one understands how the asymptotic states (here the nucleons) and the external sources (virtual photons in the present case) couple to pion pairs. 
This is the essence of a dispersive treatment for the two-pion channel, which can then be applied to the determination of electromagnetic FFs; see e.g.\ \cite{Schneider:2012ez,Hoferichter:2016duk,Alarcon:2017asr,Alarcon:2017lhg,Leupold:2017ngs}. In the present work, we are not only interested in the $Q^2$ dependence of the FFs, but in addition in their quark-mass (pion-mass) dependence. Therefore we need to know the pion-mass dependence of any input that enters our calculations. This concerns the coupling of the pions to the nucleons and to the virtual photon (pion vector FF), but also the pion p-wave phase shift. To determine the pion-mass dependence of the latter, we use the inverse amplitude method (IAM) \cite{Truong:1988zp}, which can be justified from dispersion theory \cite{Nebreda:2011di}.

To account for both the $Q^2$ and $\mpi$ dependence over a significantly broad range, 
we propose in the present paper to combine (relativistic) ChPT (including $\Delta$) with the dispersive treatment of the two-pion ($\rho$-meson) channel. In practice this implies a modification of those ChPT diagrams where the photon couples to pion pairs. 
We also show that such a dispersively modified ChPT approach is consistent with a systematic ChPT power-counting scheme. 

The previous line of reasoning has started from ChPT and added a dispersion theoretical argument. Of course, we can also start from S-matrix theory and show where we introduce ChPT as an approximation. For completeness we present also this line of arguments. 
If one writes the $S$-matrix as $S = \mathds{1} + iT$, 
then the unitarity of $S$ leads to the optical theorem \cite{Peskin:1995ev} here applied (schematically) to a nucleon FF,
\begin{eqnarray}
  \label{eq:opttheo}
  {\rm Im}T_{\gamma^* \to N  \bar N} \sim \sum\limits_i T_{\gamma^* \to i} \left( T^\dagger \right)_{i \to N \bar N} \,,
\end{eqnarray}
with the states $i$ covering all allowed hadronic intermediate states: $2\pi$, $3\pi$, $4\pi$, \ldots, $K \bar K$, $K \bar K \pi$, \ldots, 
$N \bar N$, $N \bar N \pi$, \ldots.
Note that the sum is restricted to asymptotic states:   
hadronic resonances do not appear as single-particle states 
but are accounted by the scattering amplitudes of asymptotic states. 
For isovector FFs at low energies $\sqrt{Q^2} \leq 1\,$GeV, 
the $2\pi$ intermediate state is the most important one. The basic idea of this work is therefore to treat the two-pion state via dispersion theory, relying on standard ChPT for the rest. The optical theorem (\ref{eq:opttheo}) relates the imaginary part of a loop diagram 
to the product of amplitudes. In technical terms these are the Cutkosky cutting rules.  
For ChPT diagrams this implies to relate one-loop FF diagrams to products of tree-level FF and scattering diagrams. For baryon-antibaryon intermediate states we keep just the ChPT expression, but for pion-nucleon scattering we apply modifications to the tree-level diagrams of pion-nucleon scattering and the pion vector FF. These modifications include the pion rescattering in a unitary way. One can view this as a resummation procedure of multi-loop diagrams. 

The paper is organised as follows. In subsection \ref{sec disp formalism} we will describe the dispersive formalism while in subsection \ref{sec chpt formalism} the ChPT calculation will be presented. Based on the combined formalism, we calculate the Dirac and Pauli FFs. The comparison of our results with LQCD data will be presented in section \ref{sec:F1} and section \ref{sec:F2} for Dirac and Pauli FF, respectively. Summary and outlook are provided in section \ref{sec:conclusions}. Appendices are added for technical aspects.

\section{Formalism}
\label{sec formalism}
In general, there are four electromagnetic FFs: Dirac and Pauli FFs for both proton and neutron. We write $F(q^2)$ for a generic FF, and provide labels only where it is necessary to be specific. In such cases  
we use $a=1$ (2) for the Dirac (Pauli) FF, $p$ ($n$) for the proton (neutron), and $v$ ($s$) for the isovector (isoscalar) 
combination of proton and neutron FFs, defined as
\begin{equation}
    F^{(s,v)}_a(q^2)=F^p_a(q^2)\pm F^n_a(q^2)\ .
\end{equation}
In the $q^2<0$ region the FFs are analytic functions of $q^2$; therefore, one can define mean squared radii as 
\begin{equation}
    F^{(s,v)}_a(q^2)=F^{(s,v)}_a(0)\left[1+\frac16 \, \langle r_a^{(s,v)2}\rangle \, q^2+\mO(q^4)\right] \ ,
    \label{eq:Fiexpansion}
\end{equation}
where $F_1(0)$ and $F_2(0)$ stand for the electric charge and the anomalous magnetic moment $\kappa$, respectively. 

\subsection{Dispersive Machinery}
\label{sec disp formalism}

\subsubsection{General expressions}

According to perturbative QCD \cite{Lepage:1980fj}, all FFs decrease at large $q^2$. One can then write down 
an unsubtracted dispersion relation \cite{Lin:2021umz}: 
\begin{eqnarray}
  F(q^2) = \int\limits_{s_0}^\infty \frac{ds}{\pi} \, \frac{{\rm Im}F(s)}{s-q^2 - i \epsilon} \,.
  \label{eq general disp}
\end{eqnarray}
The integral expresses the fact that the FF is an analytic function in the $q^2$ complex plane except for a cut along the real axis, which starts at the lowest threshold $s_0$ and extends to $+\infty$. Since the cut extends to the right in the standard way of displaying the complex plane (positive real part to the right), it is called ``right-hand cut''. 

In principle, the nucleon electromagnetic FFs satisfy unsubtracted dispersion relations, Eq.~(\ref{eq general disp}). The required input is the imaginary part provided by the optical theorem (\ref{eq:opttheo}). But we do not have a formula for the imaginary part that is accurate at arbitrary energies $\sqrt{s}$. ChPT works only at low enough energies. The scattering amplitudes on the right-hand side of (\ref{eq:opttheo}) are simpler the smaller the number of relevant channels. Therefore in practice also dispersion theory is typically restricted to low energies (or to an energy regime where perturbation theory in coupling constants suppresses many-particle states).  

It is common practice \cite{Schneider:2012ez,Hoferichter:2016duk,Granados:2017cib,Leupold:2017ngs,Junker:2019vvy} to use oversubtracted dispersion relations where the  sensitivity to the low-energy regime is strengthened and the sensitivity to the high-energy part is demoted. A singly-subtracted dispersion relation for the FFs is given by
\begin{eqnarray}
  F(q^2) = F(0) + q^2 \int\limits_{s_0}^\infty \frac{ds}{\pi} \, \frac{{\rm Im}F(s)}{s (s-q^2 - i \epsilon)} \,.
  \label{eq:once-subtr}
\end{eqnarray}
The additional $s$ in the denominator suppresses the high-energy part of the integrand at the price of introducing a subtraction constant $F(0)$. Recall that for the Dirac (Pauli) FF, this quantity is nothing but the electric charge (anomalous magnetic moment). 

We aim at comparing our calculation to LQCD results, where not only the momentum transfer $Q^2$ but also the quark mass is varied. In fact, the non-trivial but predictable dependence on the quark mass is a key feature of ChPT \cite{Gasser:1987rb} that we intend to maintain. Thus, in cases when the subtraction constant $F(0)$ is quark-mass dependent, we prefer the use of an unsubtracted dispersion relation (\ref{eq general disp}). Otherwise we would limit the predictive power of our expressions concerning the quark-mass dependence. However, when the subtraction constant does not depend on the quark mass, it will be of advantage to use the subtracted dispersion relation (\ref{eq:once-subtr}).

\subsubsection{The isovector channel and the $\rho$ meson}

\begin{figure}[h!]
\begin{subfigure}{0.4\textwidth}
\includegraphics[width=\textwidth]{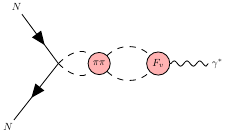}
   \end{subfigure}
   \hspace*{3em}
      \begin{subfigure}{0.4\textwidth}
\includegraphics[width=\textwidth]{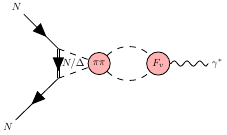}
 \end{subfigure}
\caption{Dispersively modified diagrams with a $2\pi$ cut. Solid, dashed and wiggly lines denote nucleons, pions and virtual photons. Double solid lines stand for nucleon or $\Delta$ propagators. The ``$\pi \pi$'' circle represents the pion-pion scattering S-matrix, while the ``$F_v$'' circle denotes the pion vector FF.}
\label{fig two pi cut disp}
\end{figure}

In this paper, we include the $\rho$ meson dispersively. 
As the $\rho$ appears in the isovector channel, we offer no improvement over ChPT in the isoscalar channel. Therefore we focus on the isovector FFs and from now on the superscript $(v)$ will be implicit. For $\vert q^2 \vert \leq m_\rho^2$, Eq.\,(\ref{eq general disp}) can be approximated by 
\begin{equation}
      F(q^2) \approx  \int\limits_{4\mpi^2}^{\Lambda^2} \frac{ds}{\pi} \, \frac{{\rm Im}F_{2\pi}(s)}{s-q^2 - i \epsilon}+ F_{\text{ChPT without $2\pi$ cut}}(q^2)  \,.
\label{eq disp improved form factors}
\end{equation}
The first term provides the dispersive treatment for the contribution to the FF with a two-pion cut. A subset of such contributions in which the $NN\pi\pi$ vertex is given at leading order in baryon ChPT (with explicit $\Delta$) is represented in Fig.\ \ref{fig two pi cut disp}. The second term in Eq.~(\ref{eq disp improved form factors}) accounts for all contributions without a pion cut, which we treat in perturbation theory.

In (\ref{eq disp improved form factors}) the dispersive integral is performed only up to $\sqrt{s} = \Lambda$. 
Besides the fact that our input for ${\rm Im}F_{2\pi}(s)$ (presented next) is less accurate at high $s$, the introduction of such a cut-off    
is rooted in effective field theories (and quantum field theories in general). It is a way to shift uncontrolled high-energy contributions into the counter terms (low-energy constants -- LECs). Were we only interested in a reproduction of pure ChPT, we would cut off the dispersive integral already below the $\rho$-meson mass. In ChPT this corresponds to putting the renormalisation scale to  $\Lambda_\chi \sim m_\rho$ \cite{Gasser:1983yg}. Here we aim at including the physics of the $\rho$-meson region but do not deal  with the two-baryon cut in a dispersive way. Therefore, $\Lambda$ should be chosen higher than the $\rho$-meson mass but lower than two times the nucleon mass. In this work, we set our cutoff to $\Lambda=1.8\,$GeV and come back to a possible cutoff dependence below. 

All in all, Eq.\ (\ref{eq disp improved form factors}) is the central equation for our calculation of the Dirac and Pauli FFs at low $Q^2$ and $M_\pi$.

The contribution to the FFs from the diagrams with a 2-pion cut can be calculated in dispersion theory by making use of unitarity and analyticity. Because in the isovector ($\rho$) channel, pion pairs are in a p-wave ($l=1$) state, the imaginary part of the FFs can be cast as 
\begin{eqnarray}
 {\rm Im} F_{2\pi}(s) = F_v^*(s)\, \frac{p_{\rm cm}^3 \, }{12 \pi \sqrt{s}}\,  T(s) \,.
    \label{eq imagi}
  \end{eqnarray}
The quantities $T(s)$ are the reduced $N\bar{N}\to2\pi$  p-wave amplitudes \cite{Granados:2017cib,Leupold:2017ngs}, 
%formal 
while $F_v^*$ is the conjugate of the vector FF of the pion. Finally, $p_{\rm cm} = \sqrt{s-4 \mpi^2}/2$ is the momentum of a pion in the frame where the two-pion system with invariant mass $\sqrt{s}$ is at rest; $p_{\rm cm}$ appears to the power of $2 l + 1$. Obviously, the left-hand side of 
Eq.~(\ref{eq imagi}) is real. Therefore the phases of $T$ and $F_v^*$  must cancel each other. This unitarity constraint known as Watson's theorem \cite{Watson:1954uc} prohibits the use of a purely perturbative calculation of the scattering amplitude $T$. Instead, we utilise a Muskhelishvili-Omn\`es equation \cite{zbMATH03081975,Omnes:1958hv} (see also \cite{GarciaMartin:2010cw,Kang:2013jaa,Leupold:2017ngs,Junker:2019vvy} for derivation and further discussions).
Like for the FFs themselves, we provide subtracted and unsubtracted versions for the calculation of the scattering amplitudes at low energies:
\begin{eqnarray}
  T(s) & = & K(s) + \Omega(s) \, P^{\rm unsubtr} + \Omega(s) \, 
    \int\limits_{4M_{\pi}^2}^{\Lambda^2} \, \frac{ds'}{\pi} \, 
    \frac{\sin\delta(s') \, K(s')}{\vert\Omega(s')\vert \, (s'-s-i \epsilon) }
  \label{eq:unsubtracted-T}
\end{eqnarray}
and 
\begin{eqnarray}
  T(s) & = & K(s) + \Omega(s) \, P^{\rm subtr} + \Omega(s) \, s \, 
    \int\limits_{4M_{\pi}^2}^{\Lambda^2} \, \frac{ds'}{\pi} \, 
    \frac{\sin\delta(s') \, K(s')}{\vert\Omega(s')\vert \, (s'-s-i \epsilon) \, {s'}}   \,.
  \label{eq:once subtracted T}
\end{eqnarray}
As shown below, the constant $P^{\rm unsubtr}$ can be related to LECs of ChPT and, therefore, does not have a quark-mass dependence. On the other hand, a subtracted version strengthens the contribution of the low-energy part of the integrand. Thus, also here the use of a subtracted version is advantageous provided  the corresponding constant $P^{\rm subtr}$ does not depend on the quark mass. 
In the following we omit superscript ``(un)subtr'' for $P$ whenever it is clear from the context.
In Eqs.~(\ref{eq:unsubtracted-T},\ref{eq:once subtracted T}) we have separated off the contributions with a pure left-hand cut ($K$) from the rest.\footnote{Form factors are functions of the virtuality $q^2$ alone and possess right-hand cuts in the $q^2$-plane on account of the optical theorem. Two-particle scattering amplitudes depend on the Mandelstam variables. After partial-wave projection, right-hand cuts in the variables $t$ or $u$ lead to left-hand cuts in the $s$-plane.} At tree level the scattering amplitude would just be $K(s)$ plus a polynomial in $s$. The latter is approximated by a constant $P$. 

The unitarization (or, diagrammatically,  pion-pion rescattering) responsible for the compliance of  Watson's theorem is provided by the Omn\`es function in terms of the pion-pion p-wave phase shift $\delta$:
\begin{eqnarray}
  \Omega(s) := \exp\left\{ s \, \int\limits_{4M_{\pi}^2}^\infty \frac{ds'}{\pi} \, \frac{\delta(s')}{s' \, (s'-s-i \epsilon)} \right\}\,.
  \label{eq:omnesele}  
\end{eqnarray}
At the physical pion mass, the phase shifts have been obtained and parametrized 
in Ref.~\cite{Garcia-Martin:2011iqs}. We use the IAM to obtain the light-quark mass dependence of the p-wave phase shift. Details can be found in Appendix \ref{sec alphaV}.

For the pion vector FF $F_v$ we have
\begin{equation}
\label{eq:pionFF}
F_v(s,\mpi)=[1+\alpha_V(\mpi) s]\,\Omega(s,\mpi)  \,.
\end{equation}
With the introduction of the phenomenological parameter $\alpha_V$ we follow \cite{Hanhart:2012wi,Hoferichter:2016duk,Leupold:2017ngs}. In the present work, however, we have to address in addition the pion-mass dependence of $\alpha_V$. This is also covered in Appendix \ref{sec alphaV}.

\subsection{The ChPT calculation}
\label{sec chpt formalism}
\begin{figure}[h!]
   \begin{subfigure}{0.19\textwidth}
\includegraphics[width=\textwidth]{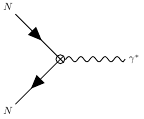}
       \caption{}
       \label{fig tree chpt}
   \end{subfigure}
   \begin{subfigure}{0.19\textwidth}
\includegraphics[width=\textwidth]{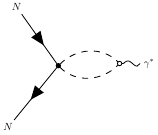}
    \caption{}
    \label{fig wt_p1}
   \end{subfigure}
      \begin{subfigure}{0.19\textwidth}
\includegraphics[width=\textwidth]{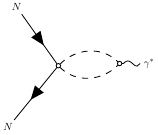}
      \caption{}
    \label{fig wt_p2}
   \end{subfigure}
    \begin{subfigure}{0.19\textwidth}
\includegraphics[width=\textwidth]{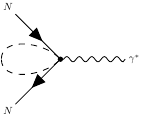}
  \caption{}
    \label{fig LO tadpole}
     \end{subfigure}
     \begin{subfigure}{0.19\textwidth}
\includegraphics[width=\textwidth]{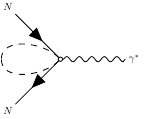}
  \caption{}
   \label{fig 2pi cut2 chpt}
   \end{subfigure}
         \begin{subfigure}{0.19\textwidth}
\includegraphics[width=\textwidth]{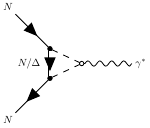}
  \caption{}
   \label{fig 1 baryon}
   \end{subfigure}
        \begin{subfigure}{0.2\textwidth}
\includegraphics[width=\textwidth]{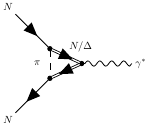}
  \caption{}
   \label{fig 2b cut 2}
   \end{subfigure}
\begin{subfigure}{0.19\textwidth}
\includegraphics[width=\textwidth]{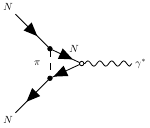}
  \caption{}
   \label{fig 2b cut}
   \end{subfigure}
            \begin{subfigure}{0.19\textwidth}
\includegraphics[width=\textwidth]{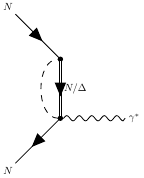}
  \caption{}
   \label{fig 6a}
   \end{subfigure}
        \begin{subfigure}{0.19\textwidth}
\includegraphics[width=\textwidth]{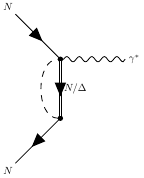}
  \caption{}
   \label{fig 6b}
   \end{subfigure}
   \caption{ChPT diagrams for the nucleon EM form factors. Line styles are defined as in Fig.\ \ref{fig two pi cut disp}. Diagram \ref{fig tree chpt} represents tree level vertices up to $\pfour$. Diagrams \ref{fig wt_p1}-\ref{fig 6b} account for $\pthree$ in a ChPT version with $\Delta$ and also $\pfour$ in a $\Delta$-less ($\slashed{\Delta}$) theory. The open and filled circles denote $\mO(p^2)$ and $\mO(p)$ vertices, respectively.}
     \label{fig chpt diagrams}
\end{figure}
We have calculated the Dirac and Pauli FFs in ChPT with explicit $\Delta$ up to $\pthree$. Additionally, $\pfour$ contributions without $\Delta$ are included for $F_2$ because we find that $\pthree$ ChPT yields unsatisfying results. On the other hand, it is not our ambition to go beyond state-of-the-art and provide a full-fledged $\pfour$ calculation that includes the $\Delta$. Therefore we make sure that we reproduce the $\pfour$ $\Delta$-less ($\slashed{\Delta}$) result of Ref~\cite{Fuchs:2003ir}.\footnote{Note that our $c_6$ LEC is $4\m$ times the one used in Ref.~\cite{Fuchs:2003ir}.} In Ref.~\cite{Bauer:2012pv} the FF is calculated up to $\pthree$ with explicit $\Delta$ and vector mesons. We reproduce the corresponding  results without vector mesons. The baryon-mass difference $\delta\equiv m_\Delta-m_N$ is counted as $\mO(p)$ (small-scale expansion \cite{Hemmert:1997ye}). This counting determines the $\pthree$ $\Delta$ loops to be included. As stated in the introduction, the calculation is relativistic and employs the EOMS renormalisation scheme \cite{Fuchs:2003qc}. This means that the power counting breaking (PCB) terms are absorbed in the LECs. In order to identify the PCB terms, one expands in $q^2$ and $\mpi$ so that the terms that violate the counting are isolated. For the identification of the PCB terms, the $\delta$ difference is not considered as an expansion parameter.

For the $\slashed{\Delta}$ terms, we take $\mL_2$ from Ref.~\cite{Gasser:1983yg} and $\mL_{\pi N}^{(1-4)}$ from Ref.~\cite{Fettes:2000gb} (we denote the chiral-limit parameters as follows: $\m$, $\md$, $\g$ and $F$).  The inclusion of the $\Delta$ is important because  certain degree of cancellation with the nucleon has been observed \cite{Leupold:2017ngs}. For the $\Delta$ loops, we employ $\mL_{\pi\Delta}^{(1)}$, $\mL_{\pi N\Delta}^{(1)}$ from Ref.~\cite{Bauer:2012pv}, but denoting the $\pi N \Delta$ coupling as $h_A$ instead of $g$. We follow the prescriptions of  Ref.~\cite{Bauer:2012pv}, setting $A=-1$. The chiral limit masses $\m=0.855$ GeV, $\md=1.166$ GeV are taken from Ref.~\cite{Alvarado:2021ibw}, and the chiral-limit value $F=0.0856$ GeV from ~\cite{FlavourLatticeAveragingGroupFLAG:2021npn}. We set $\g=g_A^{\rm phys}=1.2754$, since it is almost $\mpi$ independent \cite{Alvarado:2021ibw}, and $h_A \approx 3 \g/(2\sqrt{2}) = 1.35$. The relation between $h_A$ and $\g$ follows from QCD in the limit of  large number of colours \cite{Pascalutsa:2005nd}.\footnote{Note that in some publications $h_A$ is defined differently, e.g.\ with twice the value used here \cite{Pascalutsa:2005nd,Leupold:2017ngs}.} 

For illustrative purposes we report here the most relevant terms of the quantities defined in Eq.~\eqref{eq:Fiexpansion} (in agreement with \cite{Bauer:2012pv}): 
\bea 
    \rad^{\rm{ChPT}} &=&-12 d_6 + \frac{1}{16\pi^2F^2}\Bigg\{-2\log{\left(\frac{\mpi}{\mu}\right)}-1-\g^2\left[10 \log\left(\frac{\mpi}{\mu}\right) -12 \log \left(\frac{\m}{\mu}\right)+\frac{41}{2} \right] \no\\
    &&{}+h_A^2\bigg[\frac{379}{54}-\frac{80}{27}\log{\left(\frac{\m}{\mu}\right)}+\frac{80}{9}\log{\left(\frac{\mpi}{\mu}\right)}-\frac{80\delta \log{(X)}}{9\sqrt{\delta^2-\mpi^2}} \bigg]+ \ldots \Bigg\}\ ,
    \label{eq:r1ChPT}
\eea
with $X=(\delta-\sqrt{\delta^2-\mpi^2})/\mpi$; the ellipsis stands for higher orders in $\mpi$ and $\delta$. The full expressions are given in our supplementary material.

As already discussed, we include some $\pfour$ diagrams in our $F_2$ calculation. Here we explicitly display the full tree-level but just the leading-loop terms:
\bea
    \kappa^{\rm{ChPT}} &=& c_6-16 e_{106}\m\mpi^2+\frac{1}{4\pi^2F^2}\Bigg\{-\g^2\m\mpi +\frac{8}{9}h_A^2\m \left[\delta\log{\left(\frac{\mpi}{2\delta}\right)}-\sqrt{\delta^2-\mpi^2}\log{\left(X\right)} \right] + \ldots \Bigg\}\ ,
    \label{eq:kappaChPT}
\eea

\begin{equation}
    (\kappa\radtwo)^{\rm{ChPT}} =  12(d_6+2\m e_{74})+ \frac{1}{8\pi^2F^2}\left\{\frac{\pi\g^2\m}{\mpi}-\frac{8h_A^2\m}{9\sqrt{\delta^2-\mpi^2}}\log{\left(X\right)}+ \ldots \right\}\ .
    \label{eq:r2ChPT}
\end{equation}

\section{The Dirac form factor}
\label{sec:F1} 

The Dirac FF (for proton, isovector and isoscalar) at the photon point $q^2=0$ is given by the proton charge: 
\begin{equation}
    F_{1}(q^2=0)=1\,.
\end{equation}
This quantity is protected by gauge invariance and is therefore not renormalised. Hence it does not receive any quark-mass dependence.
Thus we can write 
\begin{eqnarray}
  &&  F_{1}(q^2) = 1 
  +  \frac{q^2}{12\pi} \, \int\limits_{4 \mpi^2}^{\Lambda^2} \frac{ds}{\pi} \, 
  \frac{T_{1}(s) \, p_{\rm cm}^3(s) \, F_v^*(s)}{s^{3/2} \, (s-q^2-i \epsilon)} +F_{1}^{\text{two-baryon cut}}(q^2)-F_{1}^{\text{two-baryon cut}}(0)+ 2q^2d_6 
  \label{eq:disp unconstrained1}  
\end{eqnarray}
where the two-pion cut is accounted by a once-subtracted dispersion relation; 
$T_{1}(s)$ is given by 
\begin{eqnarray}
  T_{1}(s) & = & K_1(s) + \Omega(s) \, P_1 + \Omega(s) \, s \, 
    \int\limits_{4\mpi^2}^{\Lambda^2} \, \frac{ds'}{\pi} \, 
    \frac{\sin\delta(s') \, K_1(s')}{\vert\Omega(s')\vert \, (s'-s-i \epsilon) \, {s'}} \,. 
  \label{eq:tmandel}
\end{eqnarray}
Diagrams \ref{fig LO tadpole}, \ref{fig 2pi cut2 chpt}, \ref{fig 6a}, \ref{fig 6b} of Fig. \ref{fig chpt diagrams} contribute only to the charge. Diagrams \ref{fig wt_p1}, \ref{fig wt_p2}, \ref{fig 1 baryon}, and parts of \ref{fig tree chpt} are covered by the dispersive integral. Diagrams \ref{fig 2b cut 2} and \ref{fig 2b cut} constitute $F_{1}^{\text{two-baryon cut}}$. These interrelations are further discussed in Appendix \ref{sec:diagr-power}.

We aim at an accuracy in the chiral counting of at least $\pthree$, i.e.\ next-to-next-to-leading order (NNLO). The Dirac FF starts at leading order (LO), but at this order one obtains only the charge but no $Q^2$ or $M_{\pi}$ dependence. At next-to-leading order (NLO) there is no new contribution \cite{Gasser:1987rb} while at NNLO all of them are proportional to $q^2$. Hence, the dispersive integral in (\ref{eq:disp unconstrained1}) requires only an LO input because the $q^2$ factor yields an overall NNLO. 
Therefore we keep in (\ref{eq:tmandel}) only the LO ChPT contribution to  
%$ T_{1}(s)$ 
$P_1$ given by 
\begin{equation}
\begin{split}
       P_1&=P^{N}_1+P^{\Delta}_1+P^{\text{WT}}_1\\
       &=-\frac{\g^2}{F^2}+\frac{2 h_A^2 (\md+\m)^2}{9 F^2 \md^2}+\frac{1}{F^2}\,
\end{split}
\label{eq:P1-formula}
\end{equation}
where $P^{N}_1$, $P^{\Delta}_1$, and $P^{WT}_1$ come from the nucleon exchange, the $\Delta$ exchange and from the Weinberg-Tomozawa term \cite{Weinberg:1966kf,Tomozawa:1966jm}, respectively.
Correspondingly, $K_1$ is obtained from the parts of the nucleon- and $\Delta$-exchange diagrams where a propagator appears (after partial-fraction decomposition) as explicitly covered in Refs. \cite{Granados:2017cib,Leupold:2017ngs}. Further details are provided in Appendix \ref{app:LHC}. 

Corrections to the LO result (\ref{eq:P1-formula}) are $\sim s$ or $\sim \mpi^2$, and therefore two powers too high. This can be most easily deduced from the results in Ref.~\cite{Gasser:1987rb} using a Ward identity that connects diagrams where one photon line is replaced by two pion lines. We use a subtracted dispersion relation in Eq.~(\ref{eq:tmandel}). The integral in this equation is nominally of higher order  but we keep it to ensure Watson's theorem.  We have checked that the once-subtracted dispersion relation (\ref{eq:disp unconstrained1}) for the Dirac FF reproduces the non-analyticity of the ChPT 2$\pi$ cut at $\pthree$.

\subsection{Comparison to LQCD results with fixed parameters}
\label{sec:F1d6fixed}

We have explored how different theories describe the LQCD results for the isovector Dirac FF $F_1$. In the present subsection we discuss results where all parameters are previously determined from experimental data. In particular, no LEC is fitted here to the LQCD results. We discuss several scenarios with and without dispersive theory  improvements. 
We test the following approaches:
\begin{itemize}
    \item A purely dispersive calculation where we neglect the contributions with a two-baryon cut and set $d_6=0$ in \eqref{eq:disp unconstrained1}. This approach is denoted by ``disp''.
    \item A plain $\pthree$ ChPT calculation without dispersive modifications [in this case the radius $\rad$ is given by Eq.~\eqref{eq:r1ChPT}]. We explore two alternatives:
    \begin{itemize}
        \item computing the amplitudes within EOMS keeping the higher order contributions to the loops, i.e. without a further expansion in small parameters. These results are labelled as ``full ChPT''.  
        \item truncating the 2$\Delta$ diagram in Fig.~\ref{fig 2b cut 2} to stay at $\pthree$. This situation is just called ``ChPT''. 
    \end{itemize}
    \item Dispersion theory supplemented with  ChPT  two-baryon loops and $\pthree$ contact term proportional to $d_6$; this is the full version of Eq. \eqref{eq:disp unconstrained1}. We consider again two alternatives:
    \begin{itemize}
        \item ``disp+full ChPT'' contains the full 2$\Delta$ loop.
        \item ``disp+ChPT'' includes the  2$\Delta$ diagram truncated up to $\pthree$.
    \end{itemize}
\end{itemize}

\begin{table}[b]
\caption{LEC $d_6$ values obtained from the experimental Dirac radius quoted by the Particle Data Group (PDG)~\cite{ParticleDataGroup:2022pth} within the various approaches (see text). 
In addition, the dependence on the renormalisation point is illustrated by showing the numbers for two typical scales. For our plots we have taken $\mu = m_\rho$.}
\centering
  \begin{tabular}{|c|c|c|c|c|}
  \hline
       & ChPT (truncated) & full ChPT & disp+ChPT (truncated) & disp+full ChPT \\
    \hline
    $d_6^{\rm exp}(\mu=m_\rho)$ (GeV$^{-2}$) & $-0.385$ & $-0.353$ & $0.216$ & $0.248$ \\
    $d_6^{\rm exp}(\mu=m_N)$ (GeV$^{-2}$)  & $-0.733$ & $-0.701$ & $-0.045$ & $-0.013$ \\
    \hline
  \end{tabular}
  \label{tab:d6exp}
\end{table}
The purely dispersive calculation, ``disp'', yields a prediction for $F_1$ without further input, whereas all the other choices depend on the LEC $d_6$. However, this parameter can be fixed from the experimental value of $\rad$ (see Table~\ref{tab:d6exp}).
We compare to the LQCD results of Ref.~\cite{Djukanovic:2021cgp}.
They are preferred over other recent LQCD determinations of the nucleon FFs such as those of Ref.~\cite{Park:2021ypf} because of the smaller dependence on lattice artifacts of the former.
Addressing these additional dependencies would complicate the analysis and is beyond the scope of the present study. 

\begin{figure}%[ht]
     \centering
     \begin{subfigure}[b]{0.46\textwidth}
         \centering
         \includegraphics[width=\textwidth]{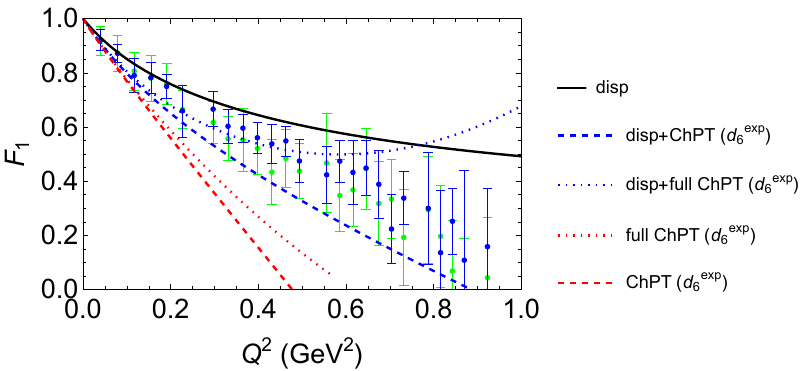}
         \caption{E250 $\mpi=0.130$ GeV}
     \end{subfigure}
     \begin{subfigure}[b]{0.46\textwidth}
         \centering
         \includegraphics[width=\textwidth]{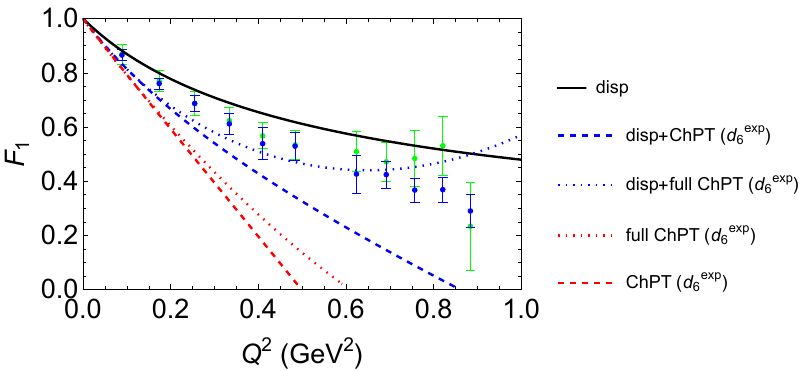}
         \caption{D200  $\mpi=0.203$ GeV}
     \end{subfigure}
     \begin{subfigure}[b]{0.46\textwidth}
         \centering
         \includegraphics[width=\textwidth]{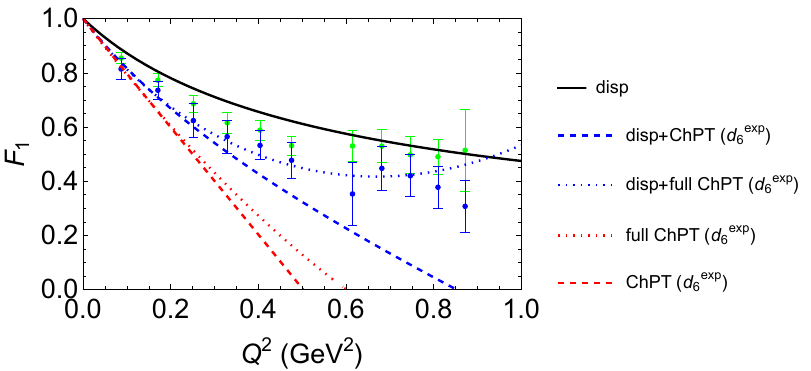}
         \caption{C101 $\mpi=0.223$ GeV}
     \end{subfigure}
     \begin{subfigure}[b]{0.46\textwidth}
         \centering
         \includegraphics[width=\textwidth]{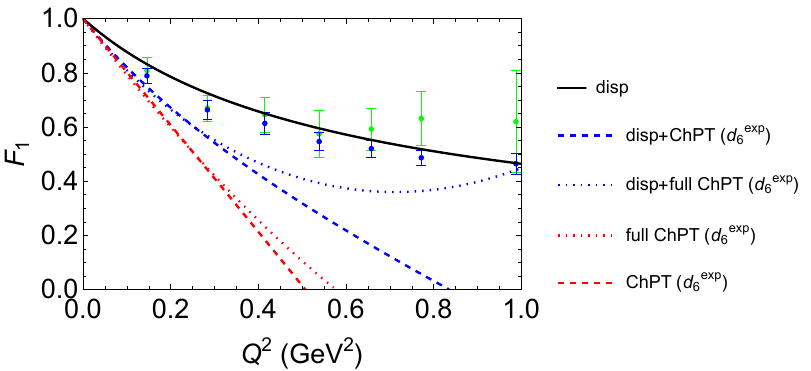}
         \caption{J303 $\mpi=0.263$ GeV}
     \end{subfigure}
     \begin{subfigure}[b]{0.46\textwidth}
         \centering
         \includegraphics[width=\textwidth]{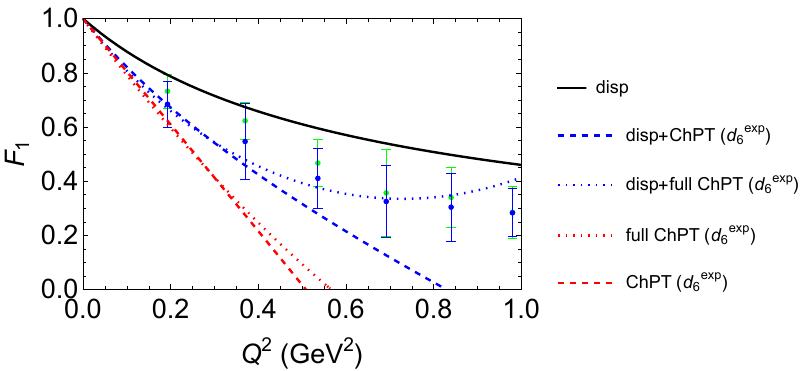}
         \caption{H105 $\mpi=0.278$ GeV}
     \end{subfigure}
     \begin{subfigure}[b]{0.46\textwidth}
         \centering
         \includegraphics[width=\textwidth]{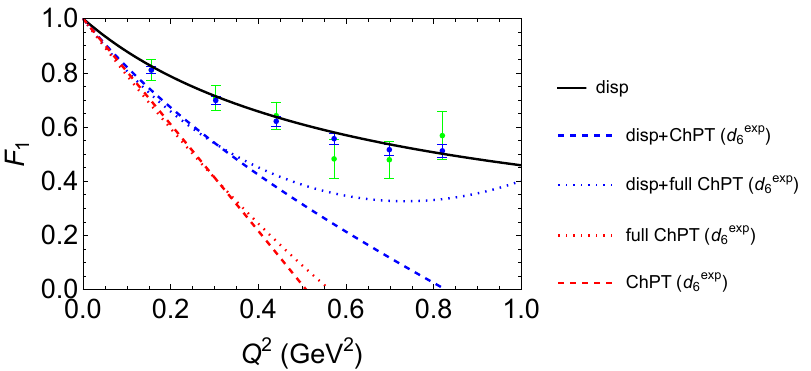}
         \caption{N200 $\mpi=0.283$ GeV}
     \end{subfigure}
     \begin{subfigure}[b]{0.46\textwidth}
         \centering
         \includegraphics[width=\textwidth]{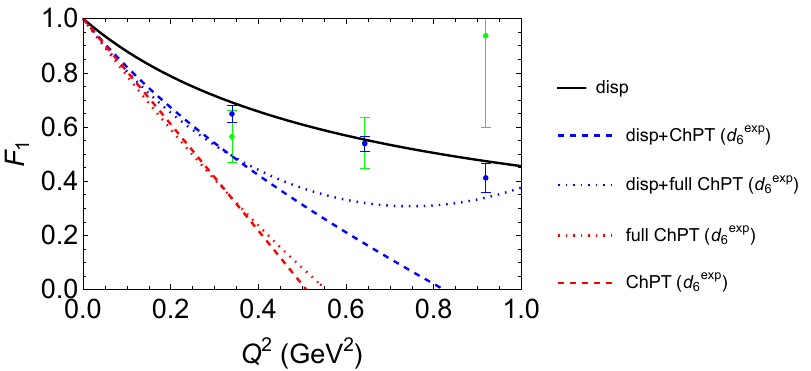}
         \caption{S201 $\mpi=0.293$ GeV}
     \end{subfigure}
    \begin{subfigure}[b]{0.46\textwidth}
         \centering
         \includegraphics[width=\textwidth]{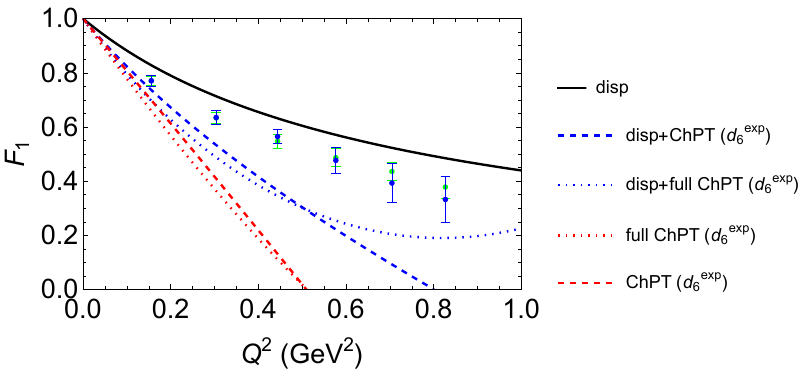}
         \caption{N203 $\mpi=0.347$ GeV}
     \end{subfigure}
     \begin{subfigure}[b]{0.46\textwidth}
         \centering
         \includegraphics[width=\textwidth]{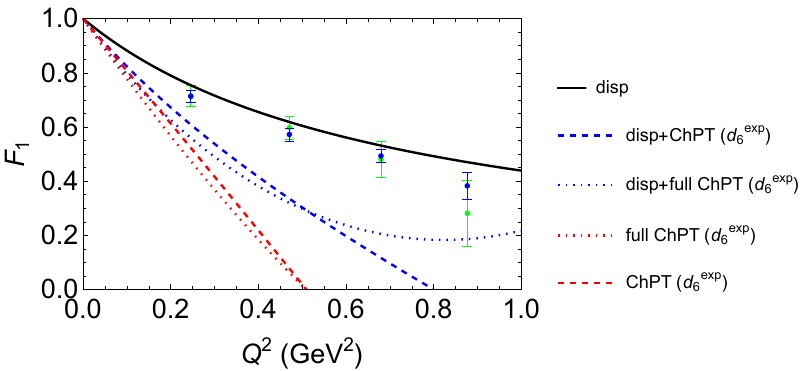}
         \caption{S400 $\mpi=0.350$ GeV}
     \end{subfigure}
    \begin{subfigure}[b]{0.46\textwidth}
         \centering
         \includegraphics[width=\textwidth]{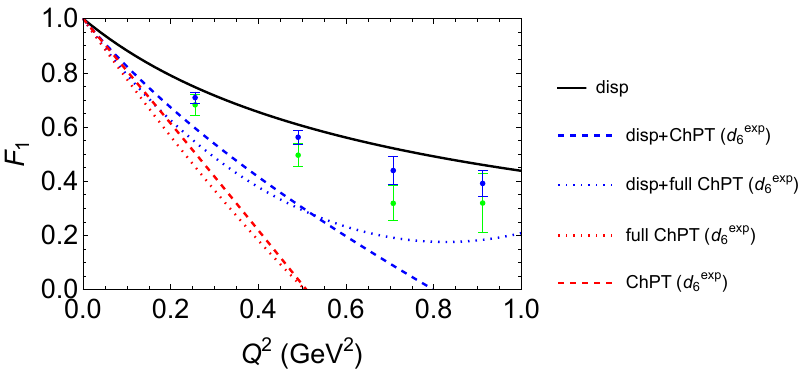}
         \caption{N302 $\mpi=0.353$ GeV}
     \end{subfigure}
 \caption{The $Q^2$ dependence of the Dirac form factor $F_1(Q^2,\mpi)$ for various pion masses, in correspondence with the LQCD ensembles of Ref.~\cite{Djukanovic:2021cgp}. The two sets of points are the LQCD results obtained with two different strategies: summation (green) and two-state method (blue); see Ref.~\cite{Djukanovic:2021cgp} for details. 
 The five curves stand for the approaches described in the text. 
 When present, the $d_6$ LEC has been extracted from the experimental value for the Dirac radius \cite{ParticleDataGroup:2022pth} (see Table~\ref{tab:d6exp}).}
          \label{fig:F1pred}
\end{figure}

\begin{figure}%[h!]
\begin{subfigure}[t]{0.45\textwidth}
%\addtocounter{subfigure}{7}
    \centering\includegraphics[width=\textwidth]{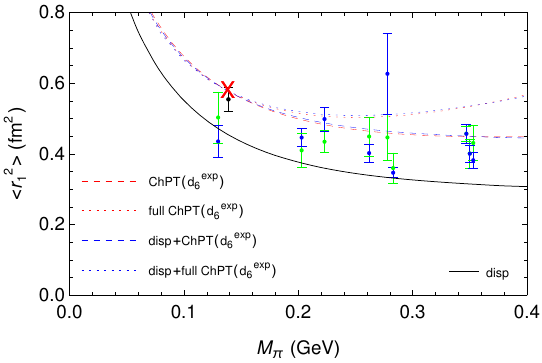}
    \caption{}
  \end{subfigure}\hspace*{3em}
  \begin{subfigure}[t]{0.45\textwidth}
    \centering\includegraphics[width=\textwidth]{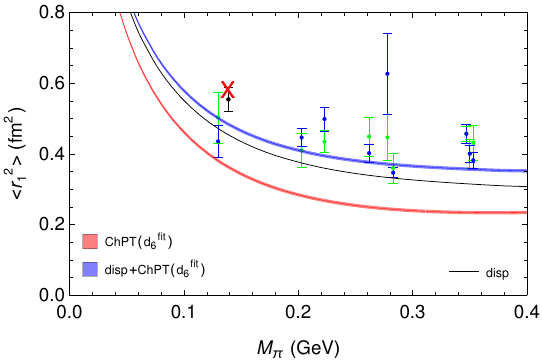}
    \caption{}
  \end{subfigure}
\caption{The Dirac radius $\rad$ as a function of the pion mass $\mpi$. 
LQCD points in green (summation method) and in blue (two-particle method) were obtained in Ref.~\cite{Djukanovic:2021cgp} using the $z$-expansion to parametrize the $Q^2$ dependence of $F_1$.  The black point is the $\rad$ value at the physical $\mpi$ obtained in Ref.~\cite{Djukanovic:2021cgp} using Heavy Baryon ChPT to extrapolate LQCD results for $F_1$ in $\mpi$ and $Q^2$. The red cross (with negligibly small error bars) corresponds to the experimental value quoted by PDG~\cite{ParticleDataGroup:2022pth}.
{\it Left panel:} results obtained with the five strategies introduced in Sec.~\ref{sec:F1d6fixed}, fixing $d_6$ at the physical $M_\pi$ with the experimental value quoted in Ref.~\cite{ParticleDataGroup:2022pth} for $\rad$. 
{\it Right panel:} results obtained by fitting $d_6$ to the LQCD values of Ref.~\cite{Djukanovic:2021cgp} for $F_1(Q^2,\mpi)$ with ``ChPT'' and ``disp+ChPT'' approaches, as discussed in the text. The bands account only for the statistical error. The ``disp'' (black) curve is the same in both panels.
}
\label{fig:r1}
\end{figure}

In Fig.~\ref{fig:F1pred} we display $F_1 (Q^2)$ at the low $M_\pi$ values of different ensembles from Ref.\ \cite{Djukanovic:2021cgp}. In a complementary plot, \ref{fig:r1}(a), the $M_\pi$ dependence of $\rad$ is shown.   
These plots show that the purely dispersive scheme (``disp'') is close to the LQCD data in both the $Q^2$ and $M_\pi$ dimensions.  The nonperturbative treatment is responsible for the generation of a realistic $Q^2$ curvature. The  $\mpi$ dependence of the radius is also well described up to $\mpi \sim 400$~MeV. In addition, the $\log(\mpi^2)$ divergence of $\rad$ at $\mpi \to 0$ predicted by ChPT is also obtained from the dispersive integral.

Turning to ChPT, one can see in Fig.\ \ref{fig:F1pred} that none of the two versions (``ChPT'' and ``full ChPT'') reproduces the $Q^2$ behaviour of $F_1$ beyond $Q^2\gtrsim 0.3\,$GeV$^2$. The resulting curvature is insufficient. We note in passing that the experimental FF also has a significant $Q^2$ curvature, as apparent from the dashed orange curve in Fig.~\ref{fig:F1fit}(a), which corresponds to the Kelly empirical  parametrization~\cite{Kelly:2004hm}. Figure \ref{fig:r1}(a) also shows that the $\mpi$ dependence of the radius is described better by the calculation with the truncated $\Delta$ contribution.

In comparison to the ChPT versions, the scheme ``disp+ChPT'' improves the $Q^2$ behaviour of the Dirac FF, though the curvature is still underestimated. In contrast, the scheme ``disp+full ChPT'' (dispersive result supplemented with untruncated ChPT) produces an excessive curvature in $Q^2$. 
In both versions, the combination of the dispersive approach and ChPT does not cause a change in the $\mpi$ dependence of $\rad(\mpi)$. Indeed, the curves for ``ChPT'' and ``disp+ChPT'' overlap in Fig.~\ref{fig:r1} and so do the ``full ChPT'' and ``disp+full ChPT'' ones. 

These comparisons show that the higher order terms present in the loop with two $\Delta$ propagators lead to a worse description of LQCD results at higher $Q^2$ and $\mpi$. Therefore, we have decided to keep from the loop with two $\Delta$ propagators only the part that is strictly ${\cal O}(p^3)$. To further support our choice, we note that the relativistic $\Delta$ propagators contain unphysical spin-1/2 contributions. In principle, those must be absorbed by LECs \cite{Pascalutsa:2000kd}. But for the two-$\Delta$ contributions beyond ${\cal O}(p^3)$, we have not written down the corresponding LECs. Eventually, a justification for our election can only come from a full-fledged calculation at ${\cal O}(p^4)$, which is beyond the scope of the present work.

\subsection{Fit to LQCD}
\label{subsec:d6}

In the previous section, by obtaining $d_6$ from the experimental value for $\rad$ we have tacitly assumed compatibility between the LQCD results and experiment. It is worthwhile to relax this constraint and attempt to fit $F_1$ with our theory, treating $d_6$ as a free parameter. On the basis of the results obtained so far we regard the dispersive calculation combined with truncated ChPT (``disp+ChPT'') as the most promising scheme for this exercise. We also fit (truncated) ``ChPT'' to have a perturbative result as a reference. 

\begin{figure}[h!]
     \centering
     \includegraphics[width=0.45\textwidth]{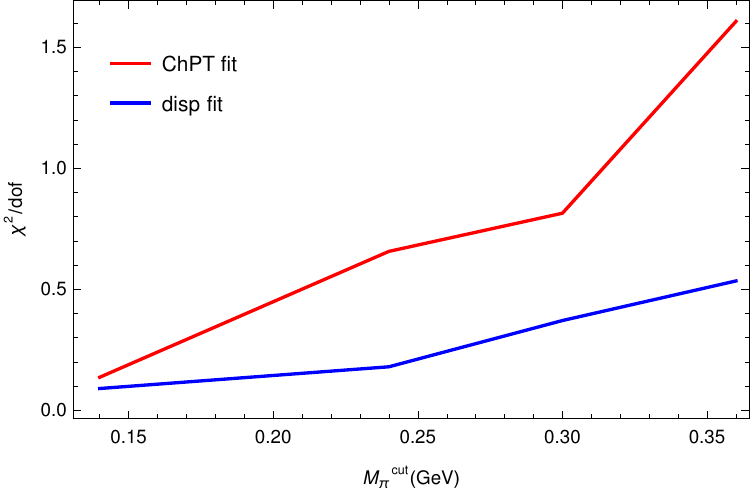}
     \caption{The value for $\chidof$ as a function of the largest pion mass $\mpi^{\rm cut}$  included in the fit. The two schemes are ``ChPT'' (red) and ``disp+ChPT'' (blue). For both we use $Q^2_{\rm cut}=0.6\,$GeV$^2$ as the largest included value.}
     \label{fig:F1chi2Mpicut}
\end{figure}

In our $\chi^2$ fits we restrict ourselves to the LQCD data sets obtained with one of the methods, the summation one, in order to avoid introducing unknown and potentially strong correlations.    
We however neglect possible correlations among different data points, keeping in mind that this approximation might cause some distortion in the interpretation of our fit.
After studying the variation of $\chi^2$ with the accepted range of $Q^2$ and $\mpi$, we find it reasonable to admit points with $Q^2<0.6\,$GeV$^2$. We include all available ensembles, so that we reach $\mpi \simeq 350\,$MeV. The evolution of $\chi^2$/dof with the variation of the largest accepted $\mpi$, while keeping the maximum $Q^2$ fixed to $0.6\,$GeV$^2$, is displayed in Fig.\ \ref{fig:F1chi2Mpicut}. 

\begin{figure}%[t!]
     \centering
          \begin{subfigure}[b]{0.3\textwidth}
         \centering
         \includegraphics[width=\textwidth]{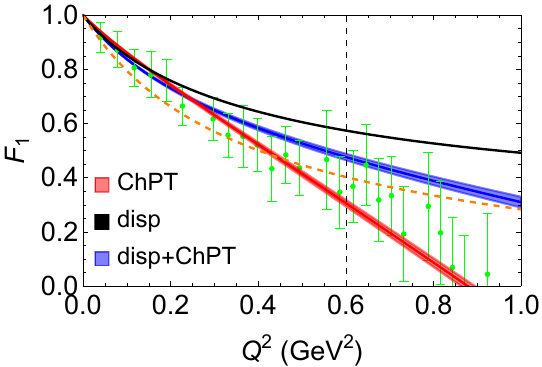}
         \caption{E250 $\mpi=0.130$ GeV}
          \label{fig:F1phys}
     \end{subfigure}
     \begin{subfigure}[b]{0.3\textwidth}
         \centering
         \includegraphics[width=\textwidth]{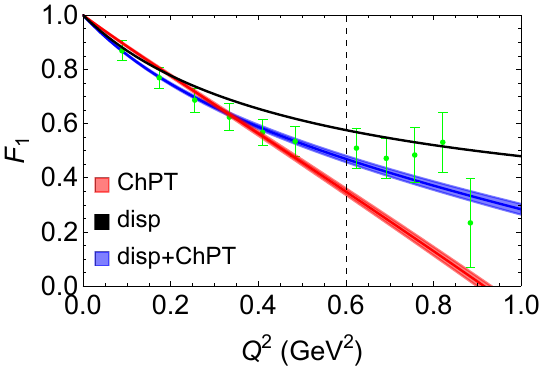}
         \caption{D200  $\mpi=0.203$ GeV}
     \end{subfigure}
     
     \begin{subfigure}[b]{0.3\textwidth}
         \centering
         \includegraphics[width=\textwidth]{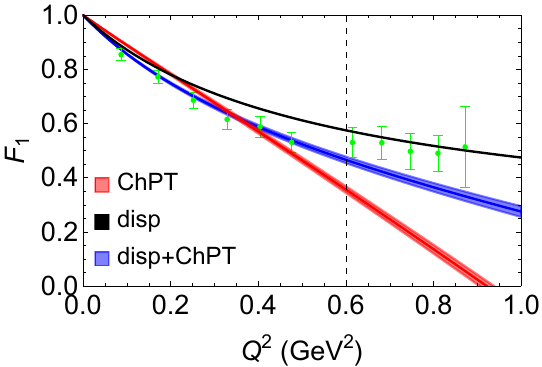}
         \caption{C101 $\mpi=0.223$ GeV}
     \end{subfigure}
     \begin{subfigure}[b]{0.3\textwidth}
         \centering
         \includegraphics[width=\textwidth]{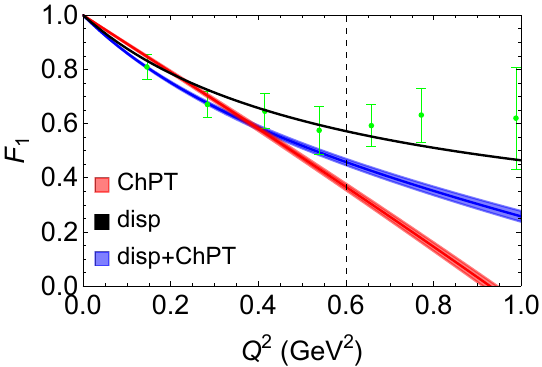}
         \caption{J303 $\mpi=0.263$ GeV}
     \end{subfigure}
     
     \begin{subfigure}[b]{0.3\textwidth}
         \centering
         \includegraphics[width=\textwidth]{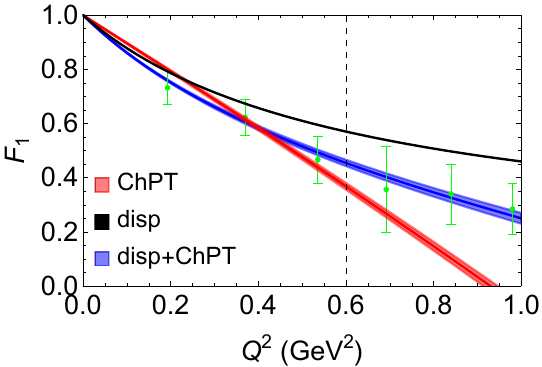}
         \caption{H105 $\mpi=0.278$ GeV}
     \end{subfigure}
     \begin{subfigure}[b]{0.3\textwidth}
         \centering
         \includegraphics[width=\textwidth]{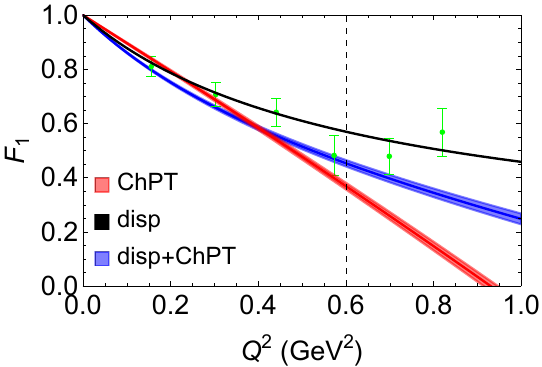}
         \caption{N200 $\mpi=0.283$ GeV}
     \end{subfigure}
     
     \begin{subfigure}[b]{0.3\textwidth}
         \centering
         \includegraphics[width=\textwidth]{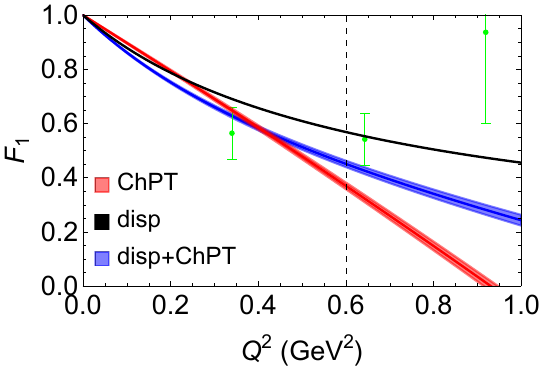}
         \caption{S201 $\mpi=0.293$ GeV}
     \end{subfigure}
    \begin{subfigure}[b]{0.3\textwidth}
         \centering
         \includegraphics[width=\textwidth]{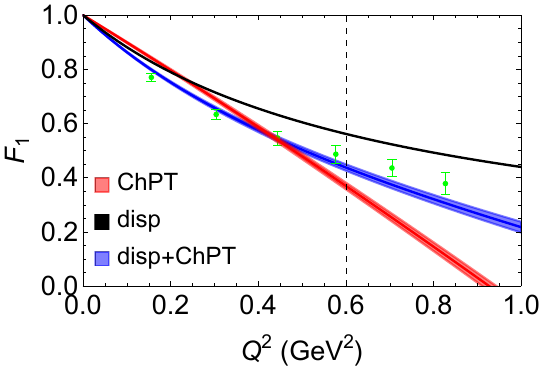}
         \caption{N203 $\mpi=0.347$ GeV}
     \end{subfigure}
     
     \begin{subfigure}[b]{0.3\textwidth}
         \centering
         \includegraphics[width=\textwidth]{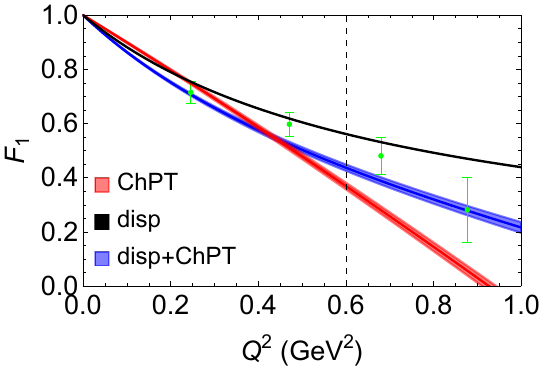}
         \caption{S400 $\mpi=0.350$ GeV}
     \end{subfigure}
     \begin{subfigure}[b]{0.3\textwidth}
         \centering
         \includegraphics[width=\textwidth]{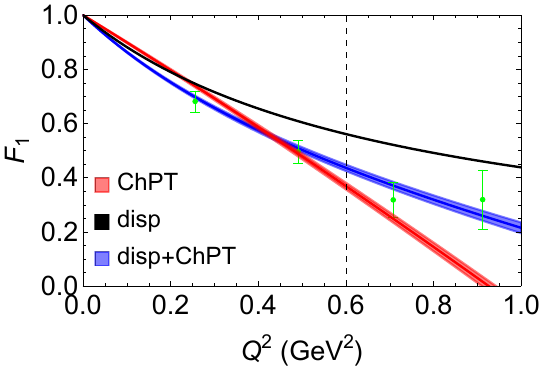}
         \caption{N302 $\mpi=0.353$ GeV}
     \end{subfigure}
 \caption{The $Q^2$ dependence of the Dirac form factor $F_1(Q^2,\mpi)$ for various pion masses, in correspondence with the LQCD ensembles of Ref.~\cite{Djukanovic:2021cgp}. Line and point styles match the ones in Fig.\ \ref{fig:F1pred} but, in contrast, the ``ChPT'' and ``disp+ChPT'' curves have been obtained by fitting the $d_6$ LEC to the LQCD points.  The vertical dashed line indicates the maximum $Q^2$ adopted in the fits. 
The dashed orange curve in panel (a) is the Kelly empirical parametrization of the $F_1$ FF~\cite{Kelly:2004hm}. }
          \label{fig:F1fit}
\end{figure}

The results of the fits using ``disp+ChPT'' and ``ChPT'' are presented in Fig.~\ref{fig:F1fit} for $F_1(Q^2,\mpi)$.  The parameter-free purely dispersive prediction, ``disp'', is also shown as a further reference. The band widths correspond to a 1$\sigma$ error in $d_6$. Note, that as an overall uncertainty in the FF determination, this error band  is underestimated, because it does not account for theoretical uncertainties arising, in particular, from the truncation of the perturbative expansion.  However, a precise determination of the error is beyond the scope of this work, and we focus on the analysis of the description of the lattice data. Table.~\ref{tab:F1fit} summarises the results of the three schemes for $d_6$, $\chi^2$, and $\rad_{\rm phys}$.
\begin{table}[h!]
\caption{Results from the $F_1(Q^2,\mpi)$ fit to LQCD.}
\centering
  \begin{tabular}{|c|c|c|c|c|c|}
  \hline
       & disp (prediction) & ChPT & disp+ChPT & HB from \cite{Djukanovic:2021cgp} &  PDG~\cite{ParticleDataGroup:2022pth} \\
    \hline
    $d_6(\mu=m_\rho)$ (GeV$^{-2}$) & - & $0.074\pm 0.010$ & $0.416\pm 0.010$ & &  \\
    $d_6(\mu=m_N)$ (GeV$^{-2}$) & - & $-0.422\pm 0.010$ & $0.155\pm 0.010$  & &  \\
    $\chidof$  & $108.9/47=2.32$ & $73.9/(47-1)=1.61$ & $24.6/(47-1)=0.53$  & & \\
    $\rad_{\rm phys}$ (fm$^2$) & 0.4541 & $0.3626\pm 0.0047$ &   $0.4838\pm 0.0047$  & $0.554\pm 0.035$ & $0.577\pm 0.0018$ \\
    \hline
  \end{tabular}
  \label{tab:F1fit}
\end{table}

By comparing Figs.\ \ref{fig:F1pred} and \ref{fig:F1fit} we observe a drastic improvement when $d_6$ is fitted to the LQCD results rather than to the experimental value of the radius. One can see in Fig.~\ref{fig:F1fit} that both ``disp+ChPT'' and ``ChPT'' are in good agreement with the LQCD data, particularly, but not only, in the $Q^2<0.6\,$GeV$^2$ region where fits were performed.
Nevertheless, focusing on the plain ``ChPT'' calculation, one has to say that the presented fit does not make the most of the theory because it is extended too high in $Q^2$.\footnote{Indeed, $Q^2\in [0.4, 0.6]$ GeV$^2$ is beyond the reach of $\pthree$ ChPT but we use the same cut as for the dispersively modified scheme to make the comparison easier.} To judge the performance of ChPT alone, we direct the reader to the previous ChPT result with $d_6^{\rm exp}(m_\rho)=-0.385$ GeV$^{-2}$ fixed to experiment; see Table\ \ref{tab:d6exp}. We actually recommend for pure ChPT to use this LEC value rather than the one fitted to the $Q^2$ dependence of the (lattice) Dirac FF. In fact, to compensate the lack of $Q^2$ curvature in ChPT, the fit yields a too small radius.

Turning now to the dispersively modified approach, one should first stress that the pure dispersive calculation is already quite good a result, as stated above. The main benefit of supplementing it with ChPT contributions is the possibility to increase the radius.  The added ChPT term mostly amounts to a shift to the radius, $\rad^{\rm disp}\to \rad^{\rm disp}-12d_6+\rad^{\text{2-baryon loops}}$ 
as can be seen in Fig.\ \ref{fig:r1}(b), (see below for a dedicated discussion).
The FF's curvature remains essentially the same. In other words, the blue (``disp+ChPT'') and black (``disp'') curves are approximately obtained from each other by rotations around the photon point. 
It is worth noticing that the $Q^2$ dependence of the LQCD results is well described with ``disp+ChPT'' up to $Q^2$ values even larger than  $Q^2_{\rm cut}=0.6\,$GeV$^2$. This scheme outperforms ``ChPT'' and ``disp'', yielding a smaller $\chi^2$ as shown in Tab.\ \ref{tab:F1fit}. In addition, at the physical $\mpi$ the ``disp+ChPT'' curve is close to the empirical Kelly  parametrization, as seen in Fig.~\ref{fig:F1fit}(a). 

The $\mpi$ dependence of $\rad$ for the fitted $d_6$ values is presented in Fig.\ \ref{fig:r1}(b). Both approaches lead to the same shape but the ``disp+ChPT'' curve is closer to the results of the extrapolations of the LQCD points to $Q^2=0$  performed in Ref.~\cite{Djukanovic:2021cgp} using the $z$ expansion. At the physical point, $\rad_{\rm phys}^{{\rm disp}+{\rm ChPT}}=0.4838\pm 0.0047\,$fm$^2$ also agrees better to the PDG value and to the heavy baryon (HB) ChPT extrapolation to the physical point of Ref.~\cite{Djukanovic:2021cgp} ($\rad^{\rm HB}=0.554\pm 0.035\,$fm$^2$) but falls short by $\sim 20$\%. 
This mismatch  could be attributed to the lack of a more realistic theoretical uncertainty in our calculation. Our results for $\rad$ and the other reference values are collected Table~\ref{tab:F1fit}.  

Ultimately, we would like to comment on the $d_6$ LEC. Its value depends on the $\mu$-running of the chiral loops. It also appears  in the disp+ChPT calculation although in this case its running comes only from the 2$\Delta$ loop. In general, $d_6$ is of the same order of magnitude in ``ChPT'' and in ``disp+ChPT''. Moreover, it stays small, $|d_6(\mu)|<2$ GeV$^{-2}$ for $\mu\in[0.5,2]$ GeV, in both approaches. The running of $d_6$ is explicitly provided in Appendix.~\ref{sec:d6}.

\section{The Pauli form factor}
\label{sec:F2} 

\subsection{Selection of diagrams}

We use an unsubtracted dispersion relation for the Pauli FF 
\begin{eqnarray}
  &&  F_{2}(q^2) =  \frac{1}{12\pi} \, \int\limits_{4 \mpi^2}^{\Lambda^2} \frac{ds}{\pi} \, 
  \frac{T_{2}(s) \, p_{\rm cm}^3(s) \, F_v^*(s)}{s^{1/2} \, (s-q^2-i \epsilon)} +F_{2}^{\text{ChPT without $2 \pi$ cut}}(q^2) 
  \label{eq:disp unconstrained2}  
\end{eqnarray}
and an unsubtracted dispersion relation for $T_2$
\begin{eqnarray}
  T_2(s) & = & K_{2}(s) + \Omega(s) \, P_{2} + \Omega(s)  \, 
    \int\limits_{4M_{\pi}^2}^{\Lambda^2} \, \frac{ds'}{\pi} \, 
    \frac{\sin\delta(s') \, K_{2}(s')}{\vert\Omega(s')\vert \, (s'-s-i \epsilon)} \,. 
  \label{eq: un subtracted T}
\end{eqnarray}
As discussed below, we need 
the nominal LO and NLO contributions to the polynomial $P_2$:
\begin{equation}
\begin{split}
P_2&=P^{N}_2+P^{\Delta}_2+P^{\text{NLO}}_2 + \mO(p^2) \\
    &= 0+\frac{4 h_A^2 \m (3 \m+4 \md)}{9 F^2 \md^2}+\frac{4 c_4 \m}{F^2} + \mO(p^2)  \, ,
\end{split}
\label{eq:P2relev}
\end{equation}
where $c_4$ is a LEC of the NLO $\pi N$ Lagrangian~\cite{Fettes:2000gb}.

In ChPT, the dominant contribution to $F_2$ appears at NLO; it is a LEC ($c_6$) which contributes to the anomalous magnetic moment in the chiral limit \cite{Gasser:1987rb}; see Eq~\eqref{eq:kappaChPT}.  Corrections at NNLO are of the form $\sim \mpi$ and $\sim Q^2/\mpi$. 
The pion mass squared scales linearly with the quark mass on account of the Gell-Mann--Oakes--Renner relation. Thus these NNLO corrections are non-analytic in the quark mass. Such terms cannot be generated by counter terms or subtraction constants. By using a subtracted dispersion relation instead of Eq.~(\ref{eq:disp unconstrained2}), one would miss part of the NNLO terms. The same logic applies to Eq.~(\ref{eq: un subtracted T}): by using a subtracted dispersion relation one would miss part of the (non-analytic) corrections, which are
 needed to achieve the required NNLO accuracy in Eq.~(\ref{eq:disp unconstrained2}). We recall that the power counting related to the dispersive integrals is discussed in Appendix \ref{sec:diagr-power}. 

On the other hand, what might be worrying is the larger cutoff sensitivity of an unsubtracted dispersion relation, Eq.~(\ref{eq:disp unconstrained2}), as compared to a subtracted one. However, if 
$s$ is large, then all quantities can be expanded in powers of $\mpi^2/s$. This part of the integration range does not generate any non-analytic behaviour in the quark mass . Therefore such cutoff dependent contributions can be compensated by counter terms. As mentioned, the Pauli FF gets a constant contribution from $c_6$, Eq.\ \eqref{eq:kappaChPT}. Additional LECs accompanied by powers of $q^2$ or $\mpi^2$ would contribute beyond the desired NNLO accuracy. 

A similar line of reasoning applies to Eq.~(\ref{eq: un subtracted T}) with its unsubtracted dispersion relation. There, the quantity $K_2$ is obtained from the tree-level nucleon- and $\Delta$-exchange diagrams of pion-nucleon scattering. One-loop diagrams with left-hand cuts lead to two-loop diagrams for the FF. This is beyond our accuracy goal. $P_2$ receives contributions from LO pion-nucleon scattering amplitudes. These are the nucleon- and $\Delta$-exchange diagrams, because they contain parts without propagators (after a partial-fraction decomposition). It turns out that, first, these contributions to $P_2$ are actually NLO; second, the nucleon contribution vanishes, and, third, the contribution from the $\Delta$-exchange depends on the details of how the $\Delta$-$N$-$\pi$ interaction term is constructed \cite{Granados:2017cib,Leupold:2017ngs}. Fortunately, there is a contact-interaction term with a LEC that appears at NLO for pion-nucleon scattering and contributes with a constant to $P_2$. This four-point interaction term, proportional to  $c_4$ in Eq.~(\ref{eq:P2relev}), absorbs the ambiguities from the three-point $\Delta$-$N$-$\pi$ interaction \cite{Pascalutsa:2000kd,Granados:2017cib,Leupold:2017ngs}. Therefore we can use $P_2$ (or $c_4$) as a fit parameter of our scheme. Besides, the dominant part of the cutoff dependence of the unsubtracted dispersion relation in Eq.~(\ref{eq: un subtracted T}) can be compensated by a change in $P_2$. 

In principle, we need LO, NLO, and NNLO terms for $T_2$ but we have already argued why a tree-level approximation for $K_2$ is sufficient. What is not covered by $K_2$ are polynomials (in $s$ and $\mpi^2$) and loop contributions without two-pion cuts. Formally at NNLO, the latter are obtained from diagrams \ref{fig LO tadpole}, \ref{fig 2b cut 2}, \ref{fig 6a}, \ref{fig 6b} in Fig.\ \ref{fig chpt diagrams} by replacing the photon line by two pion lines. Again, one can use the Ward identity in Eq. (A.10) of Ref. \cite{Gasser:1987rb} and the explicit results of Sec.~\ref{sec chpt formalism} to show that such diagrams  do not actually contribute  to $T_2$ at NNLO. 

For the polynomial $P_2$ in (\ref{eq:P2relev}), the nominal LO contribution is in practice NLO. LEC $c_4$ from the pion-nucleon NLO contact interaction contributes also with a constant. NNLO terms are only one order higher in the expansion parameter. 
Such terms cannot be analytic in $s$ or $\mpi^2$. Therefore they cannot contribute to a polynomial and we can restrict $P_2$ to a (fit) constant. Fitting $P_2$ or $c_4$ is equivalent but it might be more illuminating to use a LEC that appears in the effective Lagrangian instead of a subtraction constant of a dispersive integral.\footnote{There is no clear motivation to include other LECs in this scheme without 2-baryon loops, so we do not add them.} 
A conceptually meaningful purely dispersive approach (``disp'') starts from Eq.~\eqref{eq:disp unconstrained2}, but contains only $c_6$ instead of the full $F_{2}^{\text{ChPT without $2 \pi$ cut}}$.

On the pure ChPT side, we will find that the $\pthree$ calculation is not enough to describe the data. We display the $\pthree$ ChPT result in Fig.~\ref{fig:F2ensembles}. It predicts a too steep slope for $\kappa (\mpi)$ and a $Q^2$ dependence for $F_2$ which is incompatible with the LQCD results. For this reason, we include the $\slashed{\Delta}$ contributions of $\pfour$. Following the same criterion as for $F_1$, we truncate the $\Delta$ contribution at pure $\pthree$. For this reason, the only $\Delta$ contribution to $F_2$ comes from diagram \ref{fig 1 baryon} of Fig.\ \ref{fig chpt diagrams}. The leading contributions are given by Eqs.\ \eqref{eq:kappaChPT} and \eqref{eq:r2ChPT}.

As in the Dirac case, we combine the dispersive and ChPT contributions of the Pauli FF. 
Like for $F_1$, we add to the dispersive FF the ChPT contributions from diagrams without $2\pi$ cuts. This means that we add $\slashed{\Delta}$ diagrams \ref{fig tree chpt}, \ref{fig 2b cut 2}, \ref{fig 2b cut} and \ref{fig 2pi cut2 chpt} and the $\pthree$ $\slashed{\Delta}$ wave function renormalisation (see Tab.~\ref{tab:criterion}). For $F_2$, our truncation criterion implies that no $\Delta$ contributions are added from the ChPT side to the disp+ChPT calculation. Diagram \ref{fig 1 baryon} of Fig.\ \ref{fig chpt diagrams} is accounted by the dispersive integral. We proceed to describe how well the different parameterizations describe the LQCD data from Ref.\ \cite{Djukanovic:2021cgp}. 
\begin{table}%[h]
\caption{ChPT input for $F_2$ from the respective Feynman diagrams of Fig.\ \ref{fig chpt diagrams} that we take into account (\checkmark) or drop ($\cross$). We include all $\pthree$ and all Delta-less $\pfour$ diagrams. Therefore we exclude all $\Delta$ diagrams that de facto start at $\pfour$. $\Delta$ denotes diagrams with $\Delta$ propagators in the loop; $\slashed{\Delta}$ denotes Delta-less diagrams; ``wfr'' denotes wave-function renormalisation.}
    \begin{center}
\begin{tabular}{||c c c | c||}
 \hline
 diagrams &  ChPT & disp+ChPT & reason to include/exclude from the (disp+)ChPT scheme \\ [0.5ex] 
 \hline\hline
 \ref{fig tree chpt} & \checkmark & \checkmark & LECs\\ 
 \hline
 nucleon \ref{fig 2b cut 2}  & \checkmark & \checkmark & 2-nucleon cut diagram \\
 \hline
 \ref{fig LO tadpole} & - & - & it is zero (only contributes to $F_1(0)$) \\
 \hline
 nucleon \ref{fig 6a}, \ref{fig 6b} & - & - & it is zero  (only contributes to $F_1(0)$)  \\
 \hline
 nucleon \ref{fig 1 baryon} & \checkmark & $\cross$ & generated dispersively \\
 \hline
\ref{fig wt_p1} & - & - & it is zero  (only contributes to $F_1(0)$)  \\
 \hline 
 nucleon \ref{fig 2b cut} & \checkmark & \checkmark & $\pfour$ 2-nucleon cut diagram\\
 \hline
 \ref{fig 2pi cut2 chpt} ($c_6$) & \checkmark & \checkmark & $\slashed{\Delta}\pfour$ without cut\\
 \hline
  \ref{fig wt_p2} ($c_4$) & \checkmark &  $\cross$ & $\slashed{\Delta}\pfour$ generated dispersively \\ \hline
    $\Delta$ \ref{fig 6a}, \ref{fig 6b} & $\cross$ &  $\cross$ & de facto $\Delta\pfour$ \\ \hline
    $\Delta$ \ref{fig 2b cut 2} & $\cross$ &  $\cross$ & de facto $\Delta\pfour$ \\ \hline
    $\Delta$ \ref{fig 1 baryon} & \checkmark &  $\cross$ & generated dispersively \\ \hline
  wfr $\slashed{\Delta}\pthree\times c_6$ & \checkmark & \checkmark & de facto $\slashed{\Delta}\pfour$ without cut \\  
  \hline
  wfr $\Delta\pthree\times c_6$ & $\cross$ & $\cross$ & de facto $\pfour$ with $\Delta$ \\ 
  \hline
    wfr $\slashed{\Delta}\pfour\times c_6$ & $\cross$ & $\cross$ & de facto $\slashed{\Delta}\pfive$  \\  [1ex]
  \hline
\end{tabular}
\end{center}
    \label{tab:criterion}
\end{table}

\subsection{Fit results for $F_2$, anomalous magnetic moment and Pauli radius}

In this case, our three schemes are 
\begin{itemize}
    \item the purely dispersive approach (but including $c_6$), denoted ``disp+$c_6$''. It contains LECs $c_4$ and $c_6$ as free parameters.
    \item pure ``ChPT'' where we actually go up to $\pfour$. 
    As shown in Eqs.\ \eqref{eq:kappaChPT}, \eqref{eq:r2ChPT}, there are five LECs beyond LO ($d_6$, $c_6$, $e_{74}$, $e_{106}$ and $c_4$), but we take $d_6$ from the corresponding ``ChPT'' fit to the Dirac FF. 
    \item the combination ``disp+ChPT'', which contains the same number of fit parameters as $\pfour$ ChPT. 
\end{itemize}  
All fits are performed in the same $(Q^2,\mpi)$ region adopted for $F_1$, namely $Q^2<0.6$ GeV$^2$ and all the available $\mpi$ ensembles, i.e. $\mpi\leq 0.350\,$GeV.

\begin{figure}%[h!]
        \centering\includegraphics[width=0.45\textwidth]{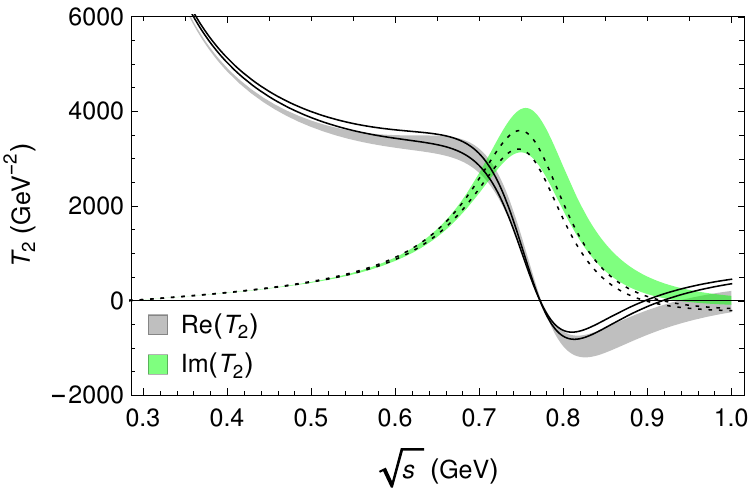}  
        \caption{Reduced amplitude $T_2$ in the unphysical region. The bands show real (gray) and imaginary (green) parts of $T_2$ as obtained from a Roy-Steiner analysis of pion-nucleon scattering \cite{Hoferichter:2015hva,Hoferichter:2016duk}. The curves represent our $T_2$, binding the region covered by the assumed prior knowledge of $c_4$. The real (imaginary) part is represented by solid (dashed) lines.}
        \label{fig:T2Roy}
\end{figure}

There is a conceptual difference between LECs $d_6$, $c_6$, $e_{74}$, $e_{106}$ on the one hand, and $c_4$ on the other. The latter is inherited from pion-nucleon scattering, while the others are directly tied to the electromagnetic FFs (tree-level contributions to magnetic moment and radii). Therefore (and in view of the relative large number of fit parameters) we constrain the fits with a Gaussian prior reflecting available knowledge about $c_4$
from pion-nucleon scattering: 
\begin{equation}\label{eq:prior}
    \chi^2=\chi_0^2+\frac{(c_4-c_4^{\rm prior})^2}{\Delta c_4^{\rm prior}} \ ,
\end{equation}
where $\chi_0^2$ denotes the standard $\chi^2$. For dispersive approaches, in order to determine this prior, we analyse the values of $c_4$ for which our reduced scattering amplitude $T_2$ agrees  well with the results obtained by solving Roy–Steiner equations for pion–nucleon scattering \cite{Hoferichter:2015hva}. This comparison is displayed in Fig.~\ref{fig:T2Roy}.
We therefore set $c_4^{\rm prior}=c_4^{\rm Roy}=-0.402$ GeV$^{-1}$ and $\Delta c_4^{\rm prior}=\Delta c_4^{\rm Roy}=0.075$ GeV$^{-1}$. We also set a prior to $c_4$ in plain ChPT. In this case we use the $\pi N$ scattering analysis of Ref.~\cite{Yao:2016vbz} and take $c_4^{\rm prior}=c_4^{\pi N}=1.200$ GeV$^{-1}$, $\Delta c_4^{\rm prior}=\Delta c_4^{\pi N}=0.045$ GeV$^{-1}$.\footnote{We use the value from \cite{Yao:2016vbz} for $c_4^{\pi N}$ even if it corresponds to a different off-shell parameter $z$. Furthermore, we neglect the $\mu$ evolution.
}

\begin{figure}%[h!]
     \centering
          \begin{subfigure}[b]{0.30\textwidth}
         \centering
         \includegraphics[width=\textwidth]{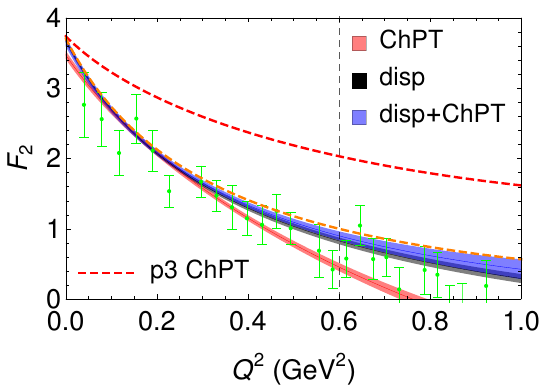}
         \caption{E250 $\mpi=0.130$ GeV}
     \end{subfigure}
     \begin{subfigure}[b]{0.30\textwidth}
         \centering
         \includegraphics[width=\textwidth]{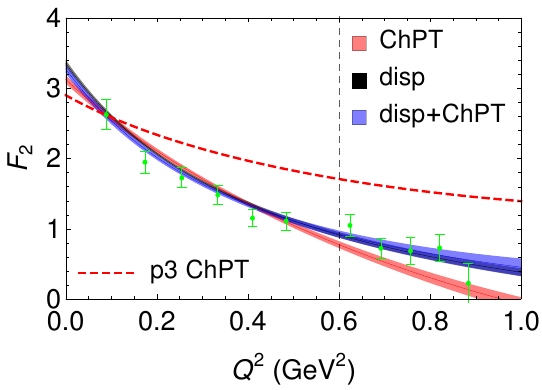}
         \caption{D200  $\mpi=0.203$ GeV}
     \end{subfigure}
     
     \begin{subfigure}[b]{0.30\textwidth}
         \centering
         \includegraphics[width=\textwidth]{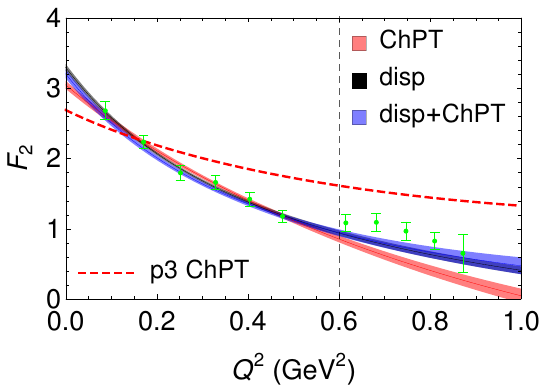}
         \caption{C101 $\mpi=0.223$ GeV}
     \end{subfigure}
     \begin{subfigure}[b]{0.30\textwidth}
         \centering
         \includegraphics[width=\textwidth]{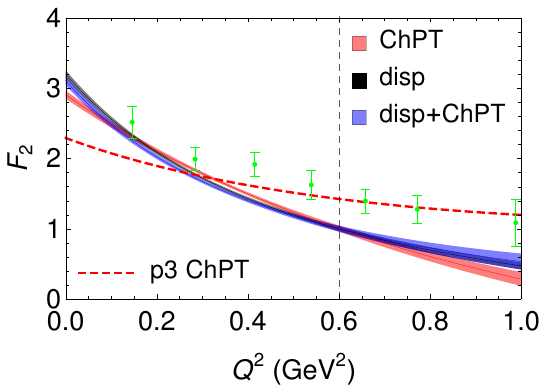}
         \caption{J303 $\mpi=0.263$ GeV}
     \end{subfigure}
     
    \begin{subfigure}[b]{0.30\textwidth}
         \centering
         \includegraphics[width=\textwidth]{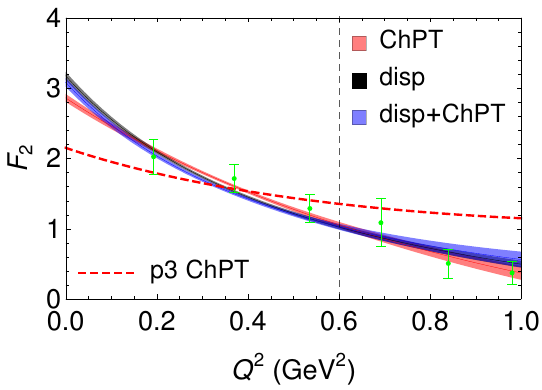}
         \caption{H105 $\mpi=0.278$ GeV}
     \end{subfigure}
      \begin{subfigure}[b]{0.30\textwidth}
         \centering
         \includegraphics[width=\textwidth]{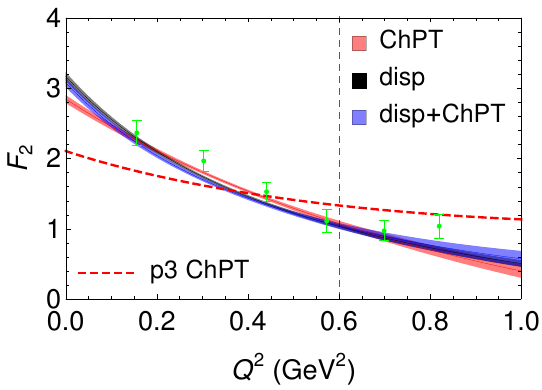}
         \caption{N200 $\mpi=0.283$ GeV}
     \end{subfigure}
     
     \begin{subfigure}[b]{0.30\textwidth}
         \centering
         \includegraphics[width=\textwidth]{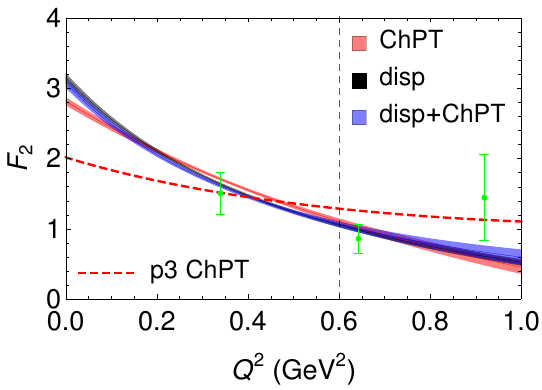}
         \caption{S201 $\mpi=0.293$ GeV}
     \end{subfigure}
     \begin{subfigure}[b]{0.30\textwidth}
         \centering
         \includegraphics[width=\textwidth]{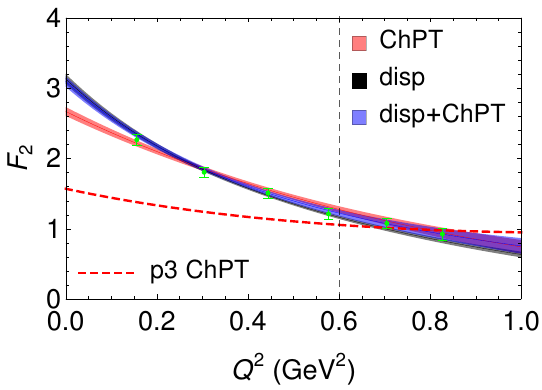}
         \caption{N203 $\mpi=0.347$ GeV}
     \end{subfigure}
     
     \begin{subfigure}[b]{0.30\textwidth}
         \centering
         \includegraphics[width=\textwidth]{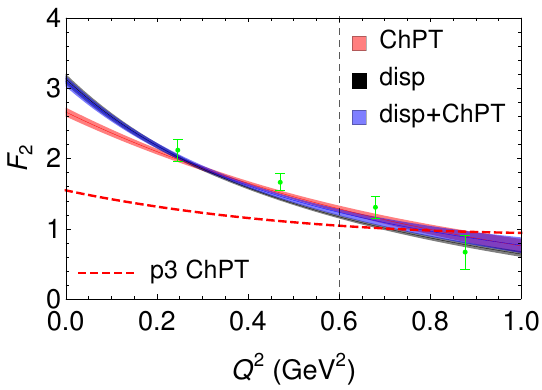}
         \caption{S400 $\mpi=0.350$ GeV}
     \end{subfigure}
     %%%%%%%%%%%%%%%%
    \begin{subfigure}[b]{0.30\textwidth}
         \centering
         \includegraphics[width=\textwidth]{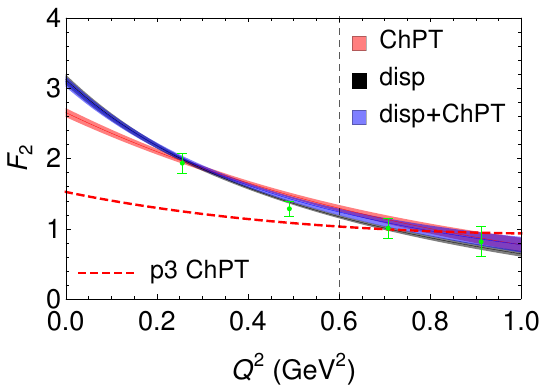}
         \caption{N302 $\mpi=0.353$ GeV}
     \end{subfigure}
     \caption{The $Q^2$ dependence of the Pauli form factor $F_2(Q^2,\mpi)$ for various pion masses, in correspondence with the LQCD ensembles of Ref.~\cite{Djukanovic:2021cgp}. LQCD points obtained with the summation method are shown. Red, black and blue bands are the results for the ``ChPT'', ``disp+$c_6$'', and ``disp+ChPT'' approaches, respectively. Band widths denote 1$\sigma$ statistical errors. The dashed red curve represents the $\pthree$ ChPT result
     The vertical dashed line indicates the maximum $Q^2$ adopted in the fits. The dashed orange curve in panel (a) is the Kelly empirical parametrization of $F_2$~\cite{Kelly:2004hm}.
     }
          \label{fig:F2ensembles}
\end{figure}
\begin{table}[h!]
        \caption{Results from our fit to the $F_2(Q^2,\mpi)$ LQCD data of Ref. \cite{Djukanovic:2021cgp}. The HB column contains the heavy-baryon extrapolation from Ref. \cite{Djukanovic:2021cgp}. The experimental values \cite{ParticleDataGroup:2022pth} are also provided.}
        \centering
          \begin{tabular}{|c|c|c|c|c|c|}
          \hline
               & disp+$c_6$ & ChPT & disp+ChPT& HB & PDG~\cite{ParticleDataGroup:2022pth} \\
            \hline
            $\chidof$   & $\frac{49.95}{47-2}=1.110$ &  $\frac{44.18}{47-4}=1.027$& $\frac{56.08}{47-4}=1.304$  & & \\
            $\chi_0^2/\rm{dof}$ &  $1.09$ & $1.027$ & $1.283$ & & \\
            $\kappa_{\rm phys}$ & $3.632\pm 0.037 $ & $3.423\pm 0.059$ & $3.605\pm 0.067$ & $3.71\pm 0.17$ & $3.706$ \\
            $\radtwo_{\rm phys}$ (fm$^2$) & $0.792\pm 0.011$ &  $0.61885\pm 0.0069$ & $0.788\pm 0.015$ & $0.690\pm 0.042$ & $0.7754\pm 0.0080$ \\
            \hline
          \end{tabular}%
          \label{tab:F2fit}
\end{table}
Let us summarise the performance of the three fits:
\begin{itemize}
    \item ``disp+$c_6$'':\\ From Fig.~\ref{fig:F2ensembles} and from the $\chi^2$ value in Table~\ref{tab:F2fit} it is apparent that dispersion theory reproduces well the $Q^2$ and $\mpi$ dependence of the LQCD data. 
    There is a large correlation between $c_4$ and $c_6$ because both of them appear in $\kappa(\mpi=0)$. 
    Actually, the LQCD data constrain $\kappa$ in the chiral limit more strongly than other quantities such as $\radtwo$. Such a mismatch among errors drives the correlation towards $-1$. 
    A fit with free $c_4$ obtains $c_4=-0.600\pm 0.031$ GeV$^{-1}$, which is close but below the Roy-Steiner value ($c_4^{\rm Roy}=-0.402\pm 0.075$ GeV$^{-1}$).
    \item ``ChPT'': 
The description of the data is good, even with better $\chi^2$ than the dispersive approaches. LEC $c_4$ goes to the prior value without causing tensions in the fit.
\item ``disp+ChPT'': 
As one can see in Fig.~\ref{fig:F2ensembles}, the best fit curve is almost identical to the ``disp+$c_6$'' one. The slight increment in $\chi^2$ compared to the other scenarios (see Table~\ref{tab:F2fit}) has no deeper meaning, because the differences are negligible. More interesting is the behaviour near $Q^2=0$, where there are no lattice points. The larger curvature of the ``disp+$c_6$'' and ``disp+ChPT'' theories make the corresponding curves steeper at $Q^2=0$ compared to ``ChPT''. This leads to the prediction of a larger radius as one can read off from Table\ \ref{tab:F2fit}. At the physical pion mass, the dispersive descriptions happen to be closer to the empirical Kelly parametrization than the  ChPT curve, describing better both the trend of the empirical curve and the LQCD points beyond the $Q^2$ cut [see Fig.~\ref{fig:F2ensembles}(a)]. We regard the combined ``disp+ChPT'' scheme as the best approach because it is more solid from the theoretical point of view if one aims at describing the FF up to rather large $Q^2 \approx 0.6\,$GeV$^2$.

\end{itemize}

\begin{table}[h!]
        \caption{Resulting values for the fitted LECs for $\mu=m_\rho$ and $\mu=m_N$ (purely dispersive scheme is $\mu$ independent).}
        \centering
          \begin{tabular}{|c|c|c|c|}
          \hline
               & disp+$c_6$ & ChPT ($\mu=m_\rho$) & disp+ChPT ($\mu=m_\rho$)\\
            \hline
            $c_4$ (GeV$^{-1}$) (with prior) & $-0.600\pm 0.031$ & $1.194\pm 0.045$ & $-0.479\pm 0.072$  \\
           $c_6$  & $-0.27\pm 0.12$  & $4.606\pm 0.057$ & $-0.88\pm 0.26$  \\
           $d_6$ (GeV$^{-2}$) (fixed) & - & $-0.385$ & $0.416$ \\
           $e_{74}$ (GeV$^{-3}$) & - & $0.178\pm 0.042$ &  $-0.293\pm 0.075$ \\
           $e_{106}$ (GeV$^{-3}$) & - & $0.170\pm 0.050$ & $-0.361\pm 0.054$  \\
           $\chidof$ & $1.110$ & $1.027$ & $1.304$ \\
           \hline
               &  & ChPT ($\mu=m_N$) & disp+ChPT ($\mu=m_N$)\\
            \hline
           $c_4$ (GeV$^{-1}$) (with prior) &  & $1.194\pm 0.045$ &  $-0.477\pm 0.072$  \\
           $c_6$  &  & $4.606\pm 0.057$ & $-0.88\pm 0.26$  \\
           $d_6$ (GeV$^{-2}$) (fixed) &  & $-0.733$ & $0.155$ \\
           
           $e_{74}$ (GeV$^{-3}$)  &  & $0.252\pm 0.042$ & $-0.140\pm 0.075$  \\
           $e_{106}$ (GeV$^{-3}$) &  & $0.151\pm 0.052$ &  $-0.4046\pm 0.060$ \\
           $\chidof$ & & $1.027$ & $1.291$ \\
            \hline
          \end{tabular}%
          \label{tab:F2fitLECs}
\end{table}
Results for the LECs are presented in Table\ \ref{tab:F2fitLECs}. 
As expected, values differ between ``ChPT'' and ``disp+ChPT''. LEC values are tied to the way how loops are renormalised. This is different in ChPT (with dimensional plus EOMS renormalisation) compared to the dispersive approach where the influence of intermediate energies is demoted by the Omn\`es function at the scale of the $\rho$-meson mass while the influence of larger energies is cut off by $\Lambda$. 
On the other hand, we note that the results do not depend strongly on $\mu$. In fact, the overall ChPT loop contribution to the radius in the ``disp+ChPT'' scheme is negligible. 
Finally, we observe that the best fit value for $c_4$ in the ``disp+ChPT'' case becomes consistent with the Roy-Steiner one: $c_4=-0.479\pm 0.072\,$GeV$^{-1}$, compared to $c_4^{\rm Roy}=-0.402\pm 0.075\,$GeV$^{-1}$. 

\begin{figure}[h!]
    \centering
\includegraphics[width=0.45\textwidth]{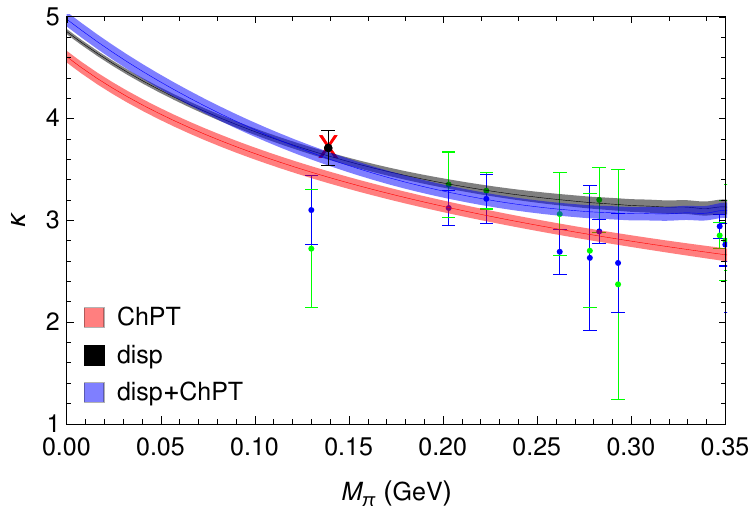} \hspace*{3em}
    \includegraphics[width=0.45\textwidth]{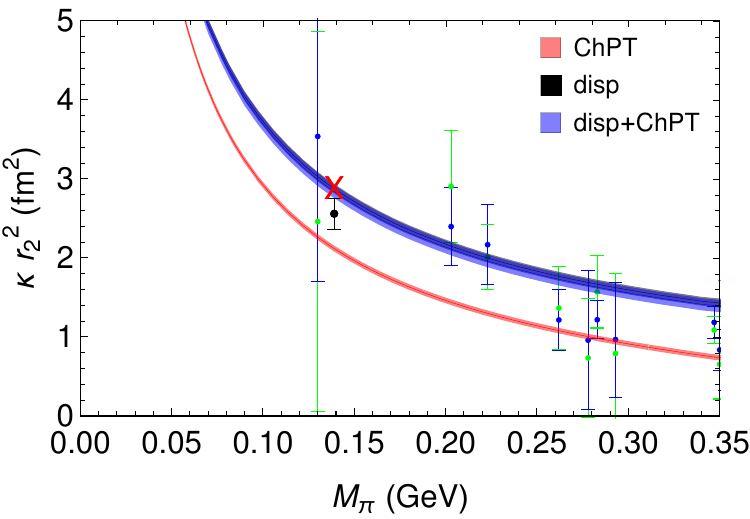}
    \caption{Pion-mass dependence of the Pauli FF and its derivative both taken at the photon point: $\kappa=F_2(0)$ (left panel) and $\kappa\radtwo = 6 F_2'(0)$ (right panel) for the three schemes. Red, black and blue bands are the results for the ``ChPT'', ``disp+$c_6$'', and ``disp+ChPT'' approaches, respectively. Band widths denote 1$\sigma$ statistical errors. LQCD points in green (summation method) and in blue (two-particle method) were obtained in Ref.~\cite{Djukanovic:2021cgp} using the $z$-expansion to parametrize the $Q^2$ dependence of $F_2$.  The black points are the values at the physical $\mpi$ obtained in Ref.~\cite{Djukanovic:2021cgp} using Heavy Baryon ChPT to extrapolate LQCD results for $F_2$ in $\mpi$ and $Q^2$. The red crosses correspond to the experimental values quoted by PDG~\cite{ParticleDataGroup:2022pth}.}
    \label{fig:kappa and r2}
\end{figure}

The $\mpi$ dependence  of $\kappa$ and $\kappa\radtwo$ according to the fits for the three approaches is given in Fig.~\ref{fig:kappa and r2}. 
The ``disp+$c_6$'' and ``disp+ChPT'' schemes yield a $\kappa(\mpi^{\rm phys})$ close to the experimental point. Our results are also in agreement with the HB extrapolation performed in Ref.~\cite{Djukanovic:2021cgp}. The ``ChPT'' curve remains slightly below the other ones. For $\radtwo$ the dispersive  results at the physical $\mpi$ are compatible with the experimental data point. The ChPT calculation is again slightly below, whereas the HB extrapolation lies in between.

\section{Conclusions}
\label{sec:conclusions} 
We have analysed the nucleon electromagnetic FFs, fundamental quantities which provide information on the nucleon structure and the underlying QCD dynamics. We calculated the isovector electromagnetic FFs combining dispersion theory and relativistic ChPT, the latter in the version with explicit $\Delta$ baryons. In particular, we  accounted for the $\mpi$ dependence of the different contributions. ChPT predicts the form and size of the non-analytic terms in $\mpi$ and these very same structures are also contained in the dispersive approach. 

In a second step, we have analysed how well we describe the LQCD data from Ref.~\cite{Djukanovic:2021cgp},  exploring three different schemes, namely (a) a purely dispersive approach that accounts only for the two-pion inelasticity as the lightest intermediate channel; (b) pure ChPT, and (c) our combined approach. For the Dirac FF, we observe that even the purely dispersive calculation is able to predict the FF reasonably well. This nonperturbative calculation provides sufficient curvature to the FF, accounting for the $\rho$-meson dynamics. We have then studied how well $\pthree$ ChPT and the combined method describe the LQCD data, fitting the $d_6$ LEC. We find that the combined dispersive and ChPT scheme outperforms the ChPT fit and the purely dispersive prediction. The calculation describes well the data for $Q^2<0.6\,$GeV$^2$ and all the $\mpi$ sampled by LQCD ($\mpi\leq 350$ MeV). We have extracted the value $\rad_{\rm phys}^{{\rm disp}+{\rm ChPT}}=0.4838\pm 0.0047$ fm$^2$ for the Dirac radius, slightly below the experimental one.

Next, we have studied the Pauli FF. Being a higher-order quantity, this required the inclusion of higher-order LECs. Therefore, we included LECs and ChPT $\slashed{\Delta}$ loops of $\pfour$. This leads to a good description of the LQCD data by both the dispersive and the ChPT calculations, in the same range of $Q^2$ and $\mpi$ as for $F_1$. Combining both theories leads essentially to the same results as the purely dispersive description. Interestingly, both the dispersive and the combined results happen to be quite close to the experimental parametrization, even beyond the $Q^2=0.6\,$GeV$^{-1}$ fit cutoff. Between these two descriptions, we regard the combined version a more solid result from the theoretical point of view.

We have extracted $\kappa_{\rm phys}^{{\rm disp}+{\rm ChPT}}=3.605\pm 0.067$, which is close to the experimental value, and $\radtwo_{\rm phys}^{{\rm disp}+{\rm ChPT}}=0.788\pm 0.015$ fm$^2$, in agreement with experiment. Furthermore, the values of several LECs have been determined, which are useful for future calculations. For this first exploration of the combined scheme, we did not attempt to determine a  theoretical uncertainty. The reported errors are purely statistical and likely to be underestimated.

To sum up, the isovector component of the Dirac and Pauli FFs are successfully described accounting not only for the $Q^2$, but also for the $\mpi$ dependence in the aforementioned range, obtaining a good agreement with lattice and experiment. We demonstrate that the dispersively modified ChPT outperforms both the dispersive method where only the $2\pi$ channel is considered and the plain ChPT without the dynamics of the $\rho$ meson. The combination of ChPT and dispersion theory improves the $Q^2$ behaviour without worsening the $\mpi$ dependence. 

There are natural extensions of the framework presented here. First of all, one can extend it to other baryon FFs as, for instance, the $\Delta$ FFs or the transition FFs from $\Delta$ to nucleon. Also for these extensions one can hope that the momentum dependence improves relative to plain ChPT while the pion-mass dependence is still properly taken care of. $\Delta$ baryons are obtained by just a spin flip of a quark in the nucleon. Therefore the properties of $\Delta$ states are tightly connected to the corresponding nucleon properties. In other words, $\Delta$ states are interesting because they provide a complementary point of view on the structure of nucleons, and are omnipresent in processes involving hadrons.  It is also conceivable to address strangeness aspects, either along the lines of  \cite{Granados:2017cib,Junker:2019vvy} where only the two-pion channel is treated dispersively, or by a full-fledged three-flavor calculation. Scientific questions that can be addressed in this way include the structure of hyperons and the strangeness content of the nucleon. 

As should be clear from our understanding of the role of Goldstone bosons as the agents of long-distance effects, the calculation of baryon low-energy properties requires a proper account of the mesonic input and the corresponding quark-mass dependence. On the other hand, mesonic properties are, of course, also interesting in their own right. In the present work we have provided as a by-product the quark-mass dependence of the pion p-wave phase shift and the pion vector FF. We propose a Blatt-Weisskopf improved IAM and an extension of the Omn\`es function that serve to include in the pion vector FF the effects beyond the two-pion channel. Such effects are small, but observable at low energies. We found good agreement with the lattice results concerning the pion-mass dependence of the mass of the $\rho$ meson. Clearly this approach might be extended to other mesons and could be further scrutinised by comparison to phase shifts and meson FFs extracted from LQCD.  

This brings us to the aspect of self-consistency given the lattice intrinsic uncertainties of continuum-limit and infinite-volume extrapolations. For an even better comparison of our calculations to lattice results it might be reasonable to use directly lattice input (instead of ChPT or IAM) for the quantities that enter our calculations (mesonic input and pion-nucleon scattering amplitudes). 

Another aspect for the very same observables, the nucleon FFs, concerns the fact that we restricted ourselves mostly to $\pthree$ calculations, at least when including the $\Delta$. Full-fledged $\pfour$ calculations in ChPT are in a development stage. In part, this relates to the excessively growing number of LECs. In addition, the role of the $\Delta$ is not so clear at this order, as we have also seen in the present work where only a restriction of the two-$\Delta$ diagram to its pure $\pthree$ part yields a reasonable curvature for the Dirac FF. On the other hand, the dispersive point of view might add some new aspects to these considerations. The $\Delta$ is an elastic pion-nucleon resonance. In this sense, the inclusion of {\em one} $\Delta$ line in a ChPT one-loop diagram can be seen as an important resummation of two- and (higher-)loop effects. Yet, the inclusion of {\em two} $\Delta$ propagators constitutes already a three-loop effect of ordinary $\Delta$-less ChPT. But how this relates to a proper power counting remains to be seen. Yet it should be clear that a reasonable $\pfour$ calculation combined with dispersion theory should help to improve the accuracy of the calculations and to provide more realistic estimates for the systematic theory uncertainties of our approach.

\begin{acknowledgments}
LAR is grateful to the Institute for Physics and Astronomy of Uppsala University for the hospitality extended during his stay. 
This work has been supported by the Swedish Research Council (Vetenskapsr\aa det) (grant number 2019-04303). It has been partially supported by the Spanish Ministerio de Ciencia e Innovaci\'on under contracts FIS2017-84038-C2-1-P and PID2020-112777GB-I00, the EU STRONG-2020 project under the program H2020-INFRAIA-2018-1, grant agreement no. 824093 and by Generalitat Valenciana under contract PROMETEO/2020/023.
\end{acknowledgments}
\clearpage
\newpage

\appendix
\section{Pion-mass dependence of mesonic quantities} 
\label{sec alphaV}

For the comparison of dispersively modified ChPT to LQCD results, it is required to know the pion-mass dependence of the pion vector FF $F_v(s,\mpi)$ and  the pion-pion (p-wave) scattering phase shifts $\delta(s,\mpi)$.

For the phase shifts, we rely on the IAM, following to some extent Ref.~\cite{Niehus:2020gmf}.
However, at next-to-leading order (NLO) the p-wave phase shifts $\delta^{\text{NLO}}_{\text{IAM}}(s)$ 
do not approach $\pi$ asymptotically as they should. 
\cite{Nebreda:2011di,Dax:2018rvs}. This problem is remedied at two-loop order by the next-to-next-to leading order (NNLO IAM) phase shifts.  
Unfortunately, at physical pion masses, the $\rho$-meson peak is not so well reproduced by the NNLO IAM fit to LQCD data \cite{Niehus:2020gmf}. For this reason, in the present work, we use NLO IAM, but instead of smoothly extrapolating the phase shift to $\pi$ \cite{Dax:2018rvs}, we modify the LO ChPT $\pi\pi$ amplitude $t_{2}(s)$ with a Blatt-Weisskopf form factor \cite{Blatt:1952ije}: 
\begin{equation}
 \Tilde{t}_2(s)=t_2(s)\frac{1}{1+r^2p_{\text{cm}}^2}=\frac{s\sigma^2}{96\pi F^2}\frac{1}{1+r^2p_{\text{cm}}^2}
 \label{eq IAM }
\end{equation}
with the velocity of the pions $\sigma(s) := \sqrt{1-4\mpi^2/s}$. 
The range parameter $r$ characterizes the scale that we do not resolve by our effective theory, i.e.\ we expect $r \sim 1/\Lambda$. 
The modified IAM amplitude ${t}^{\text{BW}}_{\text{IAM}}$ is then given as
\begin{eqnarray}
  \label{eq:IAM-test-im}
 \frac{1}{{t}^{\text{BW}}_{\text{IAM}}} =\frac{\tilde{t}_2 -\tilde{t}_4 }{\tilde{t}_2^2} = \frac{\tilde{t}_2 - \Re t_4}{\tilde{t}_2^2} - i \sigma \,,
\end{eqnarray}
where 
\begin{equation}
    \Re t_4 = \sum_{i=0}^2 b_i(s) \left[L(s)\right]^i +  \sum_{i=1}^2 b_{l_i}(s) l_i^r
\end{equation}
with $L(s)$ defined as
\begin{equation}
    L(s):=\log\frac{1+\sigma(s)}{1-\sigma(s)} \,. 
\end{equation} 
The coefficient functions are \cite{Niehus:2020gmf}
\begin{align}
b_{l_1}(s)&=-2b_{l_2}(s)=\frac{s\left(4 \mpi^2-s\right)}{48 \pi  F^4} \,, \qquad b_0(s)=
   -\frac{120\mpi^6-197 \mpi^4 s+61 \mpi^2 s^2-2 s^3}{27648 \pi ^3 F^4 \left(s-4 \mpi^2\right)},\notag\\
b_1(s)&=-\frac{64 \mpi^8-55
   \mpi^6 s+6 \mpi^4 s^2}{2304 \pi ^3 F^4 s \sigma(s)  \left(s-4 \mpi^2\right)} \,, \qquad b_2(s)=-\frac{ \mpi^4 \left(6 \mpi^4+13 \mpi^2 s-3 s^2\right)}{1536 \pi ^3 F^4 \left(s-4 \mpi^2\right)^2} \,.
\end{align}

\begin{figure}%[h!]
\centering
\begin{subfigure}[b]{0.37\textwidth}
 \centering
\includegraphics[width=\textwidth]{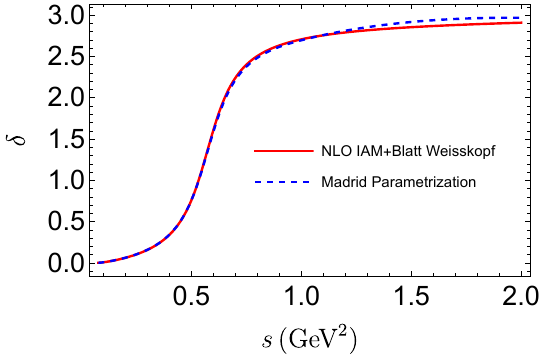}
\caption{$\delta(s)$ at the physical $\mpi$}
\label{fig: phase shifts}
 \end{subfigure}
 \begin{subfigure}[b]{0.5\textwidth}
\centering
\includegraphics[width=\textwidth]{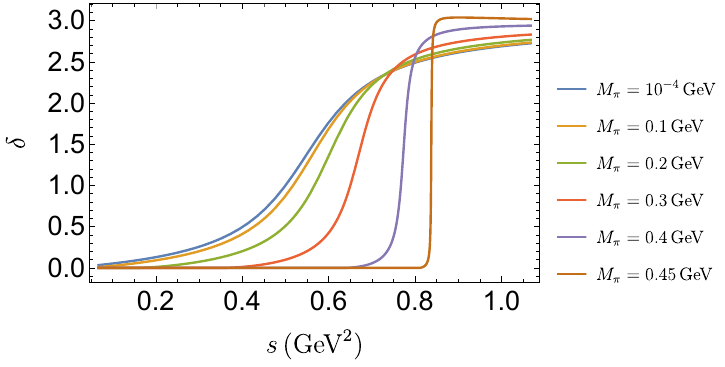}
\caption{$\delta(s)$ for different values of $\mpi$}
\label{fig: phase shifts 2}
 \end{subfigure}
 \caption{Pion p-wave scattering phase shift $\delta$ from Eq.\ (\ref{eq IAM }) as a function of the Mandelstam variable $s$.}
\end{figure}

\begin{figure}%[h!]
\centering
\begin{subfigure}[b]{0.45\textwidth}
 \centering
\includegraphics[width=0.9\textwidth]{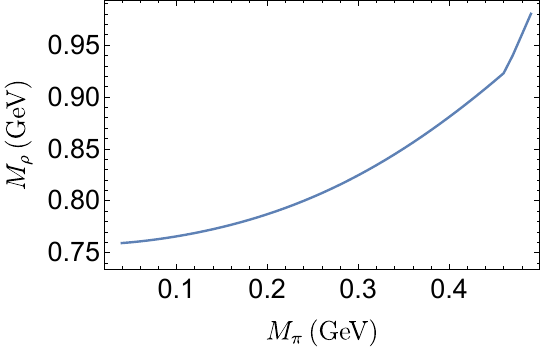}
\caption{$m_{\rho}$ as a function of $\mpi$}
\label{fig: mrho vs mpi a}
 \end{subfigure}
 \begin{subfigure}[b]{0.45\textwidth}
\centering
\includegraphics[width=0.9\textwidth]{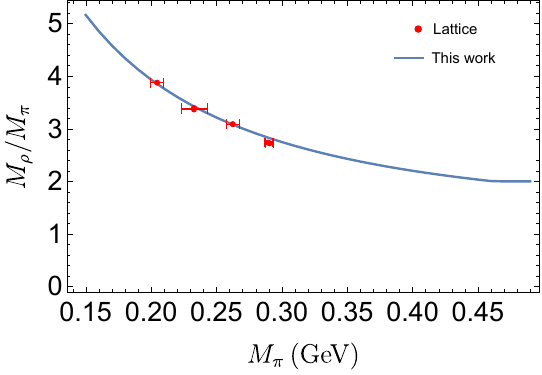}
\caption{dimensionless quantity $m_{\rho}/\mpi$ as a function of $\mpi$}
\label{fig: mrho/mpi}
 \end{subfigure}
\caption{$\mpi$ dependence of the $\rho$-meson mass $m_\rho$. The lattice data are taken from Ref.~\cite{Andersen:2018mau}.}
  \label{fig mrho vs mpi}
\end{figure}

It is easy to check that phase shifts $\delta(s,\mpi)$ extracted from ${t}^{\text{BW}}_{\text{IAM}}$ approach $\pi$ smoothly. At NLO, the combination of LECs, $l_2^r-2l^r_1$, appears in $t_4$. In Ref.~\cite{Niehus:2020gmf}, the authors find that this combination is roughly in the range $ 0.009<l_2^r-2l^r_1<0.019$. For physical pion masses, we fit $\delta(s,\mpi=0.139\, \text{GeV})$ in the range $s \in (4\mpi^2,1.5\,\text{GeV}^2)$ to the phase shifts extracted from the dispersive analysis of Ref.~\cite{Garcia-Martin:2011iqs}. We find that the best-fit values are $l_2^r-2l^r_1=0.01$ and $r=0.12\,\text{fm} = 1/(1.6\,\text{GeV})$. The resulting $\delta(s,\mpi=0.139\, \text{GeV})$ is compared to the corresponding function from Ref.~\cite{Garcia-Martin:2011iqs}. Figure\ \ref{fig: phase shifts 2} shows the phase shifts at different pion masses. It is apparent from the figure that for $M_\pi \approx 0.45$~GeV $m_{\rho} < 2 M_{\pi}$ and the $\rho$ width approaches zero. In other words,  the $\rho$-meson becomes a bound state. Above this $M_\pi$ value, the developed formalism is not directly applicable and would have to be modified to account for such a bound state.
From the crossing at $\pi/2$ we can extract the mass of the $\rho$ meson. The results are shown in Fig.\ \ref{fig mrho vs mpi}. We find that with the parameters $r$ and $l_2^r-2l^r_1$ obtained at the physical pion mass, the $\rho$ mass as a function of $M_{\pi}$ reproduces LQCD data very well \cite{Andersen:2018mau}. It is also in agreement with the three-flavor IAM results of Ref.~\cite{Molina:2020qpw}.

\begin{figure}%[h!]
    \centering
\includegraphics{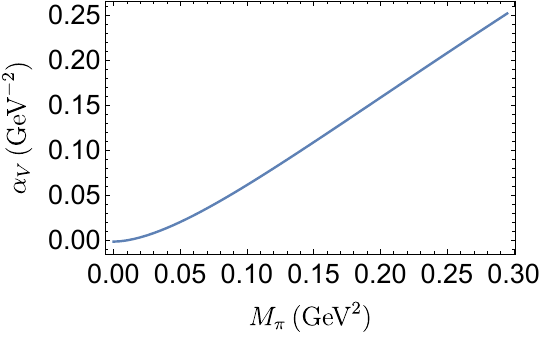}
    \caption{The dependence of the phenomenological parameter $\alpha_V$ on the pion mass.}
    \label{fig: alpha V}
\end{figure}

\begin{figure}%[h!]
    \centering
\includegraphics{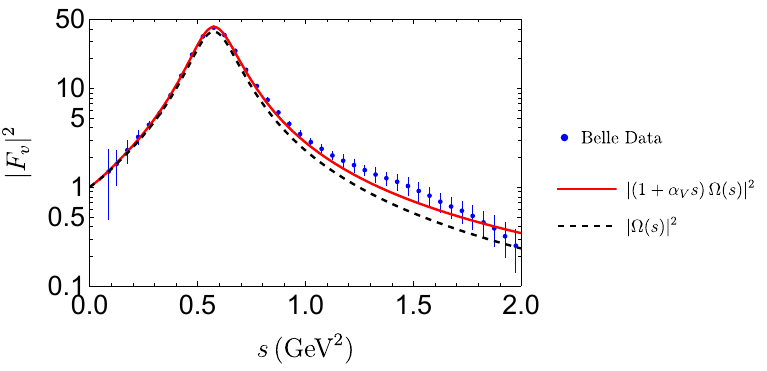}
    \caption{Our prediction for the pion vector FF $F_v(s)$ at the physical pion mass. The data are obtained from the process $\tau^- \to \pi^- \pi^0  \nu_{\tau}$ as measured by the Belle experiment \cite{Belle:2008xpe}.}
    \label{fig: pion vector FF}
\end{figure}

We turn now to the pion-vector FF, Eq.~(\ref{eq:pionFF})
\begin{equation}
F_v(s,\mpi)=[1+\alpha_V(\mpi) s]\,\Omega(s,\mpi)  \,.
\end{equation}
With the previously calculated pion phase shifts $\delta(s,\mpi)$ we can determine the Omn\`es function $\Omega(s,\mpi)$ using Eq.\ (\ref{eq:omnesele}). 
The unknown $\alpha_V(\mpi)$ has been introduced on phenomenological grounds to achieve an improved description of the experimental data for the pion vector FF \cite{Hanhart:2012wi,Hoferichter:2016duk,Leupold:2017ngs}.  
Next we determine the pion-mass dependence of $\alpha_V(\mpi)$, showing also that for physical $\mpi$ it can be predicted  from the pion charge radius. 
To tackle this issue, we start with ChPT. At NLO, the radius is \cite{Gasser:1983yg}
\begin{equation}
   \langle r_\pi^2 \rangle= \frac{1}{16\pi^2 F^2} \left(\bar l_6-1 \right)
   =: \frac{1}{16\pi^2 F^2}  \left[\tilde l_6(\mu^2) -1-\log(\mpi^2/\mu^2)\right]  \,.
\label{eq:nonumber1}
\end{equation}
To isolate the pion-mass dependence, we have introduced a LEC $\tilde l_6$ which is pion-mass independent but depends on the renormalisation scale~\cite{Gasser:1983yg}.
Using the experimental value $\langle r_\pi^2 \rangle=0.434 \, \text{fm}^2$, we find $\tilde l_6=14.26$ for $\mu = 0.770$ GeV. On the other hand, the pion charge radius is defined via the pion vector FF in the usual way, cf.\ Eq.\ \eqref{eq:Fiexpansion}. This yields 
\begin{equation}
    \alpha_V(\mpi)=\frac{\langle r_\pi^2 \rangle}{6}- \Dot{\Omega}(0,\mpi) 
    \label{eq alpha}
\end{equation}
with 
\begin{equation}
  \Dot{\Omega}(0,\mpi)= \frac{1}{\pi}\int_{4\mpi^2}^{\infty} \frac{\delta(s,\mpi)}{s^2}\,ds  \,.
  \label{eq omnes dot}
\end{equation}
Matching to $\langle r_\pi^2 \rangle$ from ChPT, Eq.\ \eqref{eq:nonumber1}, one finds the numerical $\alpha_V(\mpi)$ dependence, shown in Fig.\ \ref{fig: alpha V}. 
Note that the logarithmic pion-mass dependence in \eqref{eq:nonumber1} is compensated by a corresponding logarithm emerging from \eqref{eq omnes dot}. Therefore $\alpha_V$ has no logarithmic divergence at $\mpi =0$. As a cross-check,  in Fig.\ \ref{fig: pion vector FF} we also present the resulting $F_v(s)$ at the physical pion mass. It is displayed together with data from Belle \cite{Belle:2008xpe}. 
We observe an excellent agreement with  data up to energies of about $1\,$GeV. A pure Omn\`es function ($\alpha_V =0)$ yields a less satisfying description. 

To summarise, we have obtained reasonable parametrizations of the pion p-wave scattering phase shift and of the pion vector FF as a function of $\mpi$ and at the physical value of the latter.

\section{Diagrams generated by dispersive integrals}
\label{sec:diagr-power}

Naively, the optical theorem (\ref{eq:opttheo}) suggests that a dispersive integral produces one-loop diagrams from products of tree-level amplitudes, two-loop diagrams from products of one-loop and tree-level amplitudes and so forth. From a purely perturbative point of view (ChPT regime) this is true. However, this reasoning would not explain how the 
 whole FF (including tree level) is generated by Eq.~(\ref{eq general disp}). Therefore we need a closer look at the integration region of the dispersive integrals that we utilise. We distinguish the low-energy region of ChPT, the resonance region of the $\rho$ meson, and the high-energy region that is actually cut away by the cutoff $\Lambda$ (but might leave a $\Lambda$ dependence). Le us introduce a second cutoff $\Lambda_L$ that distinguishes the first two regions, i.e.
\begin{eqnarray}
  4 \mpi^2 \le s, s' < \Lambda_L^2 \qquad && \mbox{ChPT region,}  \nonumber \\
  \Lambda_L^2 \le s, s', m_\rho^2 < \Lambda^2 \qquad && \mbox{resonance region,}  \nonumber \\
  \Lambda^2 \le s, s'   \qquad && \mbox{high-energy region.}  
  \label{eq:regions}
\end{eqnarray}
For the following semi-quantitative discussion it is sufficient to cover the resonance region in a schematic way by viewing the $\rho$ meson as a narrow resonance. This means that in the resonance region the pion p-wave phase shift $\delta(s)$ changes rather suddenly from about 0 to $\pi$, crossing $\pi/2$ at $s=m_\rho^2$. As a consequence, we approximate 
\begin{eqnarray}
  F_v(s) \approx \Omega(s) \approx \frac{m_\rho^2}{m_\rho^2 - s - i\epsilon}
  \approx \frac{m_\rho^2}{m_\rho^2 - s - i m_\rho \Gamma_\rho}   \,, \nonumber  \\
  \Omega(s) \, F_v^*(s) \approx \vert \Omega(s) \vert^2 \approx \frac{\pi m_\rho^3}{\Gamma_\rho} \, \delta(s-m_\rho^2) 
  \label{eq:approxOmnes-VMD}  
\end{eqnarray}
where $\Gamma_\rho$  denotes the $\rho$ meson width. Finally we write the reduced scattering amplitude $T$ from Eqs.~(\ref{eq:unsubtracted-T}, \ref{eq:once subtracted T}) generically as
\begin{eqnarray}
  \label{eq:intro-R}
  T(s) =: K(s) + \Omega(s) \, R(s)  \,.
\end{eqnarray}
Obviously, this relation defines $R$ as the sum of the polynomial $P$ and the Muskhelishvili-Omn\`es integral.

These ingredients allow us to rewrite the dispersive integral in Eq.~(\ref{eq disp improved form factors}) as
\begin{eqnarray}
  \int\limits_{4\mpi^2}^{\Lambda^2} \frac{ds}{\pi} \, \frac{{\rm Im}F_{2\pi}(s)}{s-q^2 - i \epsilon}
  &\approx&
  \int\limits_{4\mpi^2}^{\Lambda^2} \frac{ds}{12\pi^2} \, \frac{K(s) \, p_{\rm cm}^3 \, \Omega^*(s)}{\sqrt{s}(s-q^2 - i \epsilon)}
            +  \int\limits_{4\mpi^2}^{\Lambda^2} \frac{ds}{12\pi^2} \, \frac{R(s) \, p_{\rm cm}^3 \, \vert \Omega(s) \vert^2}{\sqrt{s}(s-q^2 - i \epsilon)}   \nonumber \\
  &\approx&
  \int\limits_{4\mpi^2}^{\Lambda^2} \frac{ds}{12\pi^2} \, \frac{K(s) \, p_{\rm cm}^3 \, \Omega^*(s)}{\sqrt{s}(s-q^2 - i \epsilon)}
 +  \int\limits_{4\mpi^2}^{\Lambda_L^2} \frac{ds}{12\pi^2} \, \frac{R(s) \, p_{\rm cm}^3 }{\sqrt{s}(s-q^2 - i \epsilon)}
 +  \frac{m_\rho^2}{12 \pi \Gamma_\rho} \, \frac{R(m_\rho^2) \, p_{\rm cm}^3(m_\rho^2) }{m_\rho^2-q^2 - i \epsilon}  \,.
  \label{eq:split-int}  
\end{eqnarray}
We shall discuss the last three integrals one by one. For large $s$, both the left-hand-cut component $K$ and the Omn\`es function $\Omega$ decrease. Either $K$ drops so fast that the first integral is most sensitive to the ChPT region or one needs $\Omega$ to cut off the $s$-integration at
\begin{eqnarray}
  \int\limits_{4\mpi^2}^{\Lambda^2} \frac{ds}{12\pi^2} \, \frac{K(s) \, p_{\rm cm}^3 \, \Omega^*(s)}{\sqrt{s}(s-q^2 - i \epsilon)}  \,.
  \label{eq:first-K-oo3}
\end{eqnarray}
In any case, a tree-level input for $K$ is related to the one-loop diagram \ref{fig 1 baryon} of Fig.\ \ref{fig chpt diagrams}. If $K$ drops fast enough, this triangle diagram is not very sensitive to $m_\rho$ or $\Lambda$. In ChPT, the result of the integral will not depend on the renormalisation scale $\mu$. The results (dispersive and ChPT) will approximately agree. If $K$ does not drop fast enough, the dispersive expression is effectively renormalised at a scale $m_\rho$ while the ChPT diagram calculated in the standard way is renormalised at $\mu$. Thus, differences between the dispersive and  pure ChPT treatments can be compensated by readjusting counter terms (LECs), cf.\ also Appendix \ref{sec:d6}. At low values of $s$, one has $\Omega(s) \approx 1+ {\cal O}(p^2)$. If $K$ is of order ${\cal O}(p^n)$ in the chiral counting then the integral (\ref{eq:first-K-oo3}) is of order ${\cal O}(p^{n+2})$. This matches the usual expectation from ChPT: LO vertices lead to NNLO one-loop diagrams.

The integral
\begin{eqnarray}
  \int\limits_{4\mpi^2}^{\Lambda_L^2} \frac{ds}{12\pi^2} \, \frac{R(s) \, p_{\rm cm}^3 }{\sqrt{s}(s-q^2 - i \epsilon)}
  \label{eq:second-R-oo3}
\end{eqnarray}
corresponds to the ChPT diagrams \ref{fig wt_p1} and \ref{fig wt_p2} of Fig.\ \ref{fig chpt diagrams} and higher-loop diagrams. It is sensitive to the cutoff $\Lambda_L$, which again can be traded against changes of LECs.

So far we have seen that tree-level input for the scattering amplitudes leads to one-loop contributions for the FF. The less trivial aspects are related to the last term of (\ref{eq:split-int}). The fact that $\Omega$ peaks at the $\rho$-meson mass has been used to obtain
\begin{eqnarray}
  \frac{m_\rho^2}{12 \pi \Gamma_\rho} \, \frac{R(m_\rho^2) \, p_{\rm cm}^3(m_\rho^2)}{m_\rho^2-q^2 - i \epsilon}  \,.
  \label{eq:third-rho-oo3}  
\end{eqnarray}
The polynomial terms, $P$, of Eqs.~(\ref{eq:unsubtracted-T},\ref{eq:once subtracted T}) produce polynomial terms in ChPT when one expands (\ref{eq:third-rho-oo3}) in powers of $q^2/m_\rho^2$. More generally, tree-level input for the scattering amplitudes leads to tree-level contributions to the FF. One-loop input leads to one-loop contributions and so forth.

What does this mean for the power counting? In general, this means that we need an accuracy of ${\cal O}(p^n)$ in the reduced scattering amplitudes to reach an accuracy of ${\cal O}(p^n)$ for the FF. As discussed in the main text, this complication is avoided for the Dirac FF by using a subtracted dispersion relation. For the Pauli FF, however, we require input beyond LO. We actually cover a significant part of the one-loop contributions involving an NLO vertex and, in particular diagram \ref{fig wt_p2} of Fig.\ \ref{fig chpt diagrams}. To achieve a fair comparison of pure ChPT and dispersion theory, we also include ChPT $p^4$ loop diagrams. Further details are provided in Sec.~\ref{sec:F2}.

\section{The left-hand cut structures $K_{1,2}$}
\label{app:LHC}

In this Appendix, we present how to obtain from the literature the analytical expressions for the reduced amplitudes $K_{1,2}$ which enter the dispersion relation for $T_{1,2}$ in Eq.\ (\ref{eq:unsubtracted-T}).

\begin{equation}
\begin{split}
        &K_1(q^2)=\frac{8m_N^2}{4m_N^2-q^2}\left(K_E(q^2)-\frac{q^2}{4m_N^2}K_M(q^2)\right)\,,  \\
       & K_2(q^2)=\frac{8m_N^2}{q^2-4m_N^2}(K_E(q^2)-K_M(q^2)) \,.
\end{split}
\label{eq K1K2}
\end{equation}
The explicit expressions for the respective nucleon and $\Delta$ contribution to $K_{E,M}$ can be deduced from \cite{Leupold:2017ngs,Granados:2017cib}. The contact terms $P_1$ and $P_2$ are calculated in exactly the same manner as in Eq.\ (\ref{eq K1K2}).

\section{The renormalisation-group running of the LEC $d_6$}
\label{sec:d6}

The running of $d_6$ with the dimensional renormalisation scale, $\mu$, is given by
\begin{equation}
    d_6(\mu)=d_6(m_\rho)-\frac{1}{12}\frac{\beta_{d6}}{(4\pi \fpi)^2}\log\left(\frac{m_\rho}{\mu}\right) \,,
\end{equation}
where in ChPT $\beta_{d6}=\beta_{d6}^{\text{no}\Delta\Delta}+\beta_{d6}^{\Delta\Delta}$, while in disp+ChPT the running is given just by the loop with two $\Delta$, $\beta_{d6}=\beta_{d6}^{\Delta\Delta}$. The beta functions read
\begin{equation}
    \beta_{d6}^{\text{no}\Delta\Delta}= 2-2 \g^2+4 h_A^2 \frac{\left(\m^2+2 \m \md-8 \md^2\right)}{9 \md^2}, \quad \beta_{d6}^{\Delta\Delta}=10 h_A^2 \frac{\left(7 \m^4+6 \m^3 \md+9 \m^2 \md^2+16 \m \md^3-48 \md^4\right)}{27 \md^4} \ .
\end{equation}
The reason why the running differs in the two schemes can be traced back to the fact that the dispersively treated loop diagrams do not require an explicit renormalisation. Instead, the Omn\`es function cuts off the integrals at around the $\rho$-meson mass. 

\clearpage 
\newpage
\bibliography{main}

\end{document}